\documentclass[12pt, draftclsnofoot, journal, letterpaper, onecolumn]{IEEEtran}

\makeatletter
\def\ps@headings{%
\def\@oddhead{\mbox{}\scriptsize\rightmark \hfil \thepage}%
\def\@evenhead{\scriptsize\thepage \hfil \leftmark\mbox{}}%
\def\@oddfoot{}%
\def\@evenfoot{}}
\makeatother 

\IEEEoverridecommandlockouts

\usepackage{amsfonts}
\usepackage[dvips]{graphicx}
\usepackage{caption}
\usepackage{subfigure}
\usepackage{times}
\usepackage{cite}
\usepackage{lettrine}
\usepackage{amsmath}
\usepackage{amsmath}
\usepackage{amsthm}
\allowdisplaybreaks[4]
\usepackage{array}
\usepackage{amssymb}

\usepackage{stfloats}
\usepackage{slashbox}
\usepackage{graphicx}
\usepackage{footnote}
\usepackage{booktabs}
\usepackage{array}
\usepackage{algorithmic}
\usepackage{algorithm}
\usepackage{subeqnarray}
\usepackage{cases}
\usepackage{threeparttable}
\usepackage{color}

\begin{document}

\title{\LARGE{Offloading and Resource Allocation with General Task Graph in Mobile Edge Computing: A Deep Reinforcement Learning Approach} }

\IEEEoverridecommandlockouts
\author{Jia~Yan,~\IEEEmembership{Student Member,~IEEE}, Suzhi~Bi,~\IEEEmembership{Senior Member,~IEEE}, and Ying-Jun~Angela~Zhang,~\IEEEmembership{Fellow,~IEEE}
\thanks{
Part of the work has been submitted to the IEEE International Conference on Communications (ICC), Dublin, Ireland, Jun. 7-11, 2020 \cite{myICC}.
J. Yan (yj117@ie.cuhk.edu.hk) and Y. J. Zhang (yjzhang@ie.cuhk.edu.hk) are with the Department of Information Engineering, The Chinese University of Hong Kong, Hong Kong. They are also with the Shenzhen Institute of Artificial Intelligence and Robotics for Society (AIRS). S. Bi (bsz@szu.edu.cn) is with the College of Electronic and Information Engineering, Shenzhen University, Shenzhen, China.}
}

\maketitle

\vspace{-1.5cm}

\begin{abstract}
In this paper, we consider a mobile-edge computing (MEC) system, where an access point (AP) assists a mobile device (MD) to execute an application consisting of multiple tasks following a general task call graph. The objective is to jointly determine the offloading decision of each task and the resource allocation (e.g., CPU computing power) under time-varying wireless fading channels and stochastic edge computing capability, so that the energy-time cost (ETC) of the MD is minimized. Solving the problem is particularly hard due to the combinatorial offloading decisions and the strong coupling among task executions under the general dependency model. Conventional numerical optimization methods are inefficient to solve such a problem, especially when the problem size is large.  To address the issue, we propose a deep reinforcement learning (DRL) framework based on the actor-critic learning structure. In particular, the actor network utilizes a DNN to learn the optimal mapping from the input states (i.e., wireless channel gains and edge CPU frequency) to the binary offloading decision of each task. Meanwhile, by analyzing the structure of the optimal solution, we derive a low-complexity algorithm for the critic network to quickly evaluate the ETC performance of the offloading decisions output by the actor network. With the low-complexity critic network, we can quickly select the best offloading action and subsequently store the state-action pair in an experience replay memory as the training dataset to continuously improve the action generation DNN.  To further reduce the complexity, we show that the optimal offloading decision exhibits an one-climb structure, which can be utilized to significantly reduce the search space of action generation.   Numerical results show that for various types of task graphs, the proposed algorithm achieves up to $99.1\%$ of the optimal performance while significantly reducing the computational complexity compared to the existing optimization methods.

\end{abstract}

%
%
%

\section{Introduction}

Recent years have witnessed explosive growth of Internet of Things (IoT) as a way to connect tens of billions of resource-limited wireless devices, such as sensors, mobile devices (MDs) and wearable devices, to Internet through the cellular networks. Due to small physical sizes and stringent production costs constraints, IoT devices often suffer from limited computation capabilities and finite battery lives. Perceived as a promising solution, mobile edge computing (MEC) \cite{MECsurvey1,MECsurvey2} has attracted significant attention. With MEC, computationally intensive tasks can be offloaded to nearby servers located at the edges of wireless networks. This efficiently overcomes the drawbacks of long backhaul latency and high overhead compared to traditional mobile cloud computing.

Typically, there are two computation task offloading models for MEC \cite{MECsurvey1}: one is referred to as binary offloading, and the other is partial offloading. For the binary offloading model, each task is either executed locally or offloaded to the MEC server as a whole \cite{MEC3,xu,MEC2,MEC5,MEC7,MEC4}. As for partial offloading, tasks can be arbitrarily divided into two parts that are executed by the device and the edge server, respectively \cite{partial1,partial2}. Nevertheless, in practice, a mobile application usually has multiple components and the dependency among them cannot be ignored since the outputs of some components are the inputs of others. In this regard, task call graph \cite{cs_taskgraph} is proposed to model the sophisticated inter-dependency among different components in a mobile application. In this paper, we consider computation offloading with a general task call graph.

Due to the random variation of wireless channels, it is not always advantageous to offload all the tasks for edge execution. Instead, offloading computation tasks in an opportunistic manner considering the time-varying channel condition has shown significant performance advantage \cite{MEC3,xu,MEC2,MEC5,MEC7,MEC4,partial1,partial2}. Due to the mutual coupling constraints in a task call graph, offloading policy design becomes much challenging \cite{single4,single5,single6,single7,multi1,mypaper}. Specifically, \cite{single4} considered a sequential task graph and derived an optimal one-climb policy, where the execution migrates only at most once between the MD and the cloud server. This work was extended to a general task graph case in \cite{single5}, where authors applied the partial critical path analysis for the general task graph scheduling.  In \cite{single6}, the offloading problem in a general task graph was formulated as a linear programming problem through convex relaxation.  \cite{single7} modeled the task scheduling problem in a general task graph as an energy consumption minimization problem that is solved by a genetic algorithm. Note that general task graphs are considered much harder to deal with compared to other task graphs with special structures (i.e., sequential task graph), since it is hard to explore and derive the offloading properties (i.e., one-climb policy in the sequential task graph) with the general and complicated coupling among tasks.

On the other hand, recent work has considered joint optimization of radio/computing resource allocation and computation offloading. In particular, \cite{multi1} studied an energy-efficiency cost minimization problem by incorporating CPU frequency control and transmit power allocation in the MEC offloading decision.  \cite{mypaper} considered  inter-user task dependency and proposed a reduced-complexity Gibbs sampling algorithm to obtain the optimal offloading decisions.


The existing work on task offloading with general task graph adopts either convex relaxation methods (e.g., in \cite{multi1,single6}) or heuristic local search methods (e.g., in \cite{single4,single5,single7,mypaper}). However, both methods are likely to get stuck in a local optimal solution that does not guarantee good performance. Moreover, the optimization problems need to be re-solved once the wireless channel conditions change or the available computing power of the edge server changes due to the variation of demands by background applications. The frequent re-calculation of offloading decisions renders the existing methods impractical.

In this paper, we endeavor to design an efficient optimal computation offloading algorithm in an MEC system with a general task graph, so that the optimal decision swiftly adapts to the time-varying wireless channels and available edge computing power with very low computational complexity. In particular, we propose a deep reinforcement learning (DRL) framework. The key idea of DRL is to utilize the deep neural networks (DNNs) to learn the optimal mapping between the state space and the action space.  There exists several work on DRL-based offloading methods for MEC systems \cite{DRL2,DRL1,DRL4}. In \cite{DRL2}, a deep Q-network (DQN) based offloading policy was proposed to optimize the computational performance in the MEC system with energy harvesting. When tasks arrive randomly, \cite{DRL1} proposed DQN to learn the optimal offloading decisions without a priori knowledge of network dynamics. To tackle the curse of dimensionality problem in DQN-based methods, \cite{DRL4} proposed a novel DRL framework to achieve near-optimal offloading actions by considering only a small subset of candidate offloading actions in each iteration. Notice that \cite{DRL2,DRL1,DRL4} all assume independent tasks among multiple users.  Very recently, considering a general task dependency, \cite{DRL5} proposed a recurrent neural network (RNN) based reinforcement learning method for the computation offloading problem. However, it neglected the system dynamics, such as wireless fading channels and time-varying edge server CPU frequency.

We consider an MEC system with a single access point (AP) and a MD as shown in Fig. 1. The MD has an application with a general task topology to execute under time-varying wireless fading channels and edge server CPU frequency. In particular, we propose a DRL framework to minimize the weighted sum of task execution time and energy consumption of the MD. The main contributions are concluded as follows:
\begin{itemize}
  \item We formulate a mixed integer optimization problem to jointly optimize the offloading decisions and local CPU frequencies of the MD to minimize the computation delay and energy consumption. The problem is challenging because of the combinatorial nature of the offloading decisions and the strong coupling among task executions under general dependency model.

  \item In order to solve the combinatorial optimization problem efficiently, we propose a DRL  framework based on the actor-critic learning structure, where we train a DNN in the actor network periodically from the past experiences to learn the optimal mapping between the states (i.e., wireless channels and edge CPU frequency) and actions (i.e., offloading decisions). Within the actor network, we devise a novel Gaussian noise-added order-preserving action generation method to balance the diversity and complexity in generating candidate binary offloading actions under a high-dimensional action space.

  \item For the critic network, we simplify the problem according to the total loop-free paths in the general task graph and derive closed-form solution for the optimal local CPU frequencies. Based on this, we propose an efficient algorithm. As such, unlike traditional actor-critic networks that utilize a DNN to predict the values of the actions in the critic network, our analysis allows fast and accurate calculation of the performance of each action generated by the actor network. In this way, the complexity and convergence of the actor-critic based DRL are greatly improved.
  \item To further speed up the computation of the proposed DRL framework,  we propose a heuristics where the offloading decisions are limited to the ones that follow the one-climb offloading policy.  The heuristics greatly reduces the number of performance evaluations for the actions in the critic network. The optimality of the one-climb policy is analyzed and its advantageous performance over conventional action generation method is verified through simulations.

\end{itemize}

Numerical results show that for various types of general task graphs, the proposed DRL-based algorithm achieves up to $99.1\%$ of the optimal energy and time cost. Meanwhile, our proposed method only takes around 1 second to generate an offloading action, which is more than one order of magnitude faster than the other representative benchmark methods. {\color{blue}{In this paper, we formulate the joint optimization of offloading and resource allocation with general task graph in the MEC as a mixed integer non-linear programming (MINLP) problem, which is hard to solve with conventional optimization algorithms under time-varying wireless channels and stochastic edge computing capability. By exploring the special structure of the considered MINLP problem, we observe that for any given integer variables (offloading decisions), the remaining problem is convex. Therefore, the main difficulty lies in finding the optimal integer offloading decisions. With such property, we propose the actor-critic learning structure based DRL algorithm, where the actor network generates a set of integer offloading actions according to the time-varying parameters and the critic network scores each action output from the actor network by convex optimization. Then, we utilize the generated action-score pairs to make current offloading decision and improve the performance of the actor network. It is worth mentioning that the key target of the critic is for evaluating the action quality, regardless of using a general neural network or a specialized algorithm \cite{add1}. In this paper, as one of the major contributions, we propose an efficient low-complexity algorithm in the critic network to evaluate the actions generated from the actor network, which greatly reduces the training cost of the critic DNN and increases the accuracy of action evaluation.}}

The rest of the paper is organized as follows. In Section II, we present the system model and problem formulation. The optimal local CPU frequencies under fixed offloading decisions are studied in Section III.  We introduce the detailed design for the DRL framework in Section IV. In Section V, simulation results are described. Finally, we conclude the paper in Section VI.



\section{System Model And Problem Formulation}

As shown in Fig. 1, we consider an MEC system with one AP and one MD. The AP is the gateway of the edge cloud and has stable power supply. The MD has a computationally intensive mobile application consisting of $M$ dependent tasks. The input-output dependency of the tasks is represented by a directed acyclic task graph $G=(\mathcal{M},\mathcal{E})$. As shown in Fig. 2, each vertex  in $G$ represents a task $i$ and the associated parameter $L_i$ indicates the computing workload in terms of the total number of CPU cycles required for accomplishing the task. Besides, each edge $(k,i)\in \mathcal{E}$ in $G$ represents that a precedent task $k$ must be completed before starting to execute task $i$. Additionally, we denote the size of data in bits transferred from task $k$ to $i$ by $O_{k,i}$. For simplicity of exposition, we introduce two virtual tasks $0$ and $M+1$ as the entry and exit tasks, respectively. Specifically, we have $L_0=L_{M+1}=0$. By forcing the two virtual tasks to be executed locally, we ensure that the application is initiated and terminated at the MD side. We denote the set of tasks in the task graph $G$ as $\mathcal{M}=\{0,1,...,M+1\}$.

\begin{figure*}[htbp]
\centering
\begin{minipage}[t]{0.48\textwidth}
\centering
\includegraphics[scale=0.8]{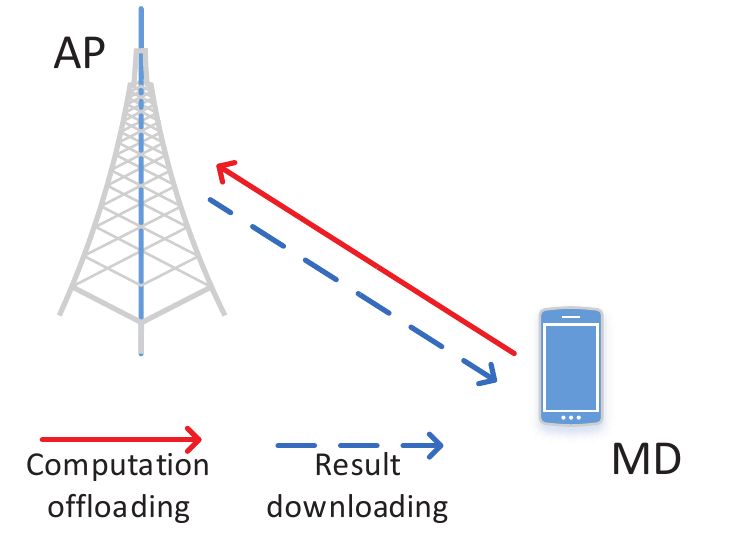}
\caption{System model.}
\end{minipage}
\begin{minipage}[t]{0.48\textwidth}
\centering
\includegraphics[scale=0.5]{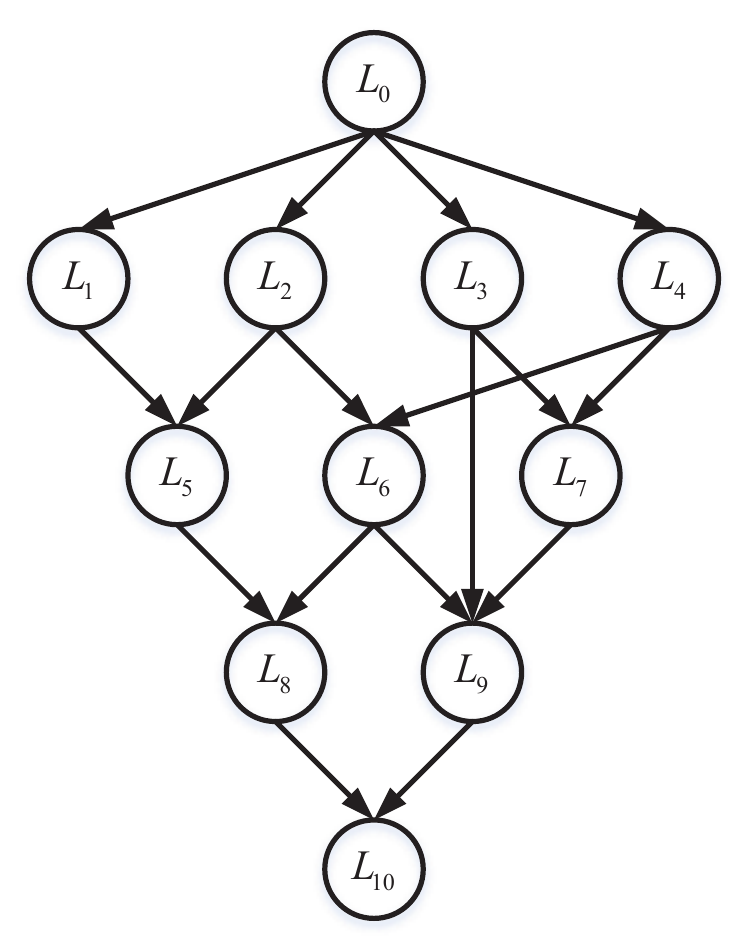}
\caption{The considered task graph.}
\end{minipage}
\end{figure*}

%

Define an indicator variable $a_{i}\in\{0,1\}$ such that $a_{i}=0$ means that task $i$ is executed locally and $a_{i}=1$ means that the MD offloads the computation of  task $i$ to the edge side. Recall that the two virtual tasks $0$ and $M+1$ must be executed locally. That is, $a_0=a_{M+1}=0$.

In addition, we assume that the MD is allocated a dedicated spectral resource block throughout its transmission, which can support concurrent transmissions for task offloading and downloading. {\color{blue}{We denote by $h_{k,i}^u$ and $h_{k,i}^d$ the channel gains when offloading and downloading the task data $O_{k,i}$, respectively.}}
Besides, we assume additive white Gaussian noise (AWGN) with zero mean and equal variance $\sigma^{2}$ at the receiver for all the tasks.

To characterize the task execution time and energy
consumption for local and edge computing, respectively, we first define the \emph{finish time} and \emph{ready time} of each task.

\emph{\textbf{Definition 1} (Finish Time).} The finish time of task $i$ is the moment when all the workload $L_i$ has been executed. We denote $FT_{i}^{l}$ and $FT_{i}^{c}$ as the finish time of task $i$ when it is executed locally and at the edge server, respectively.

\emph{\textbf{Definition 2} (Ready Time).} The ready time of a task is the earliest time when the task has received all the necessary input data to commence the task computation. For instance, in Fig. 2, the ready time of the fifth task is the time when both the input data streams from the first and second tasks have arrived. We denote the ready time of task $i$ when computing locally and at the edge server as $RT_{i}^{l}$ and $RT_{i}^{c}$, respectively.

\subsection{Local Computing}

We assume that the MD is equipped with a $\rho^{l}$-core CPU, where each CPU core can execute only one task at a time. That is, the MD can execute in total $\rho^l$ tasks simultaneously. Suppose that task $i$ is computed locally. We denote the local CPU frequency for computing the task as $f_{i}^{l}$, which is upper bounded by $f_i^l\leq f_{peak}$. Thus, the local execution time of task $i$ is given by
\begin{align}\label{f_mapping}
\tau_{i}^{l}=\frac{L_{i}}{f_{i}^{l}},
\end{align}
and the corresponding energy consumption is \cite{MECsurvey1}
\begin{align}\label{local_energy}
e_{i}^{l}=\kappa L_{i}(f_{i}^{l})^{2}=\kappa\frac{L_{i}^3}{(\tau_{i}^{l})^2},
\end{align}
where $\kappa$ is the effective switched capacitance depending on the chip architecture. {\color{blue}{According to the circuit theory \cite{add3}, the power consumption of the CPU is approximately proportional to the product of $V_{cir}^2f_i^l$, where $V_{cir}$ is the circuit supplied voltage.  Besides, $V_{cir}$ is approximately linear proportional to the CPU frequency $f_i^l$ when the CPU works at the low voltage limits \cite{add2}. Therefore, the energy consumption per CPU cycle is given by $\kappa(f_i^l)^2$.}} {\color{blue}{It is worth mentioning that for the two virtual tasks $0$ and $M+1$, we have $\tau_{0}^l=\tau_{M+1}^l=0$ and $e_{0}^l=e_{M+1}^l=0$.}}

If a task $k$ preceding task $i$ is executed at the edge server, then the output data $O_{k,i}$ must be downloaded to the MD before task $i$ can be executed locally. Denote the fixed downlink transmit power of the AP by $P_{AP}$. Then, {\color{blue}{according to the Shannon-Hartley theorem,}} the downlink data rate from the AP to the MD is
\begin{align}\label{downlink_rate}
R_{k,i}^{d}=W\log_{2}\left(1+\frac{P_{AP}h_{k,i}^d}{\sigma^{2}}\right).
\end{align}
The corresponding downlink transmission time for sending the data $O_{k,i}, (k,i)\in\mathcal{E},$ is
\begin{align}
\tau_{k,i}^{d}=\frac{O_{k,i}}{R_{k,i}^{d}}.
\end{align}

As such, the ready time $RT_{i}^{l}$ of task $i$ is given by
\begin{align}\label{RT_local}
RT_{i}^{l}=\max_{k\in\textbf{pred(i)}}\left\{(1-a_{k})FT_{k}^{l}+a_{k}\left(FT_{k}^{c}+\tau_{k,i}^{d}\right)\right\},
\end{align}
where $\textbf{pred(i)}$ denotes the set of immediate predecessors of task $i$. Specifically, if $a_k=1$ for a task $k\in\textbf{pred(i)}$, the time until its output data is available at the MD for the execution of task $i$ is equal to its finish time $FT_{k}^{c}$ at the edge side plus the downlink transmission time $\tau_{k,i}^{d}$. Otherwise, if $a_k=0$, the time until its output data is available at the MD is equal to its local finish time $FT_{k}^{l}$. When all needed data is available at the ready time $RT_i^l$, the MD locally computes task $i$ with the local execution time $\tau_i^l$ in \eqref{f_mapping}, so that the finish time of task $i$ becomes
\begin{align}\label{FT_local}
FT_{i}^{l}=RT_{i}^{l}+\tau_{i}^{l}.
\end{align}

\subsection{Edge Computing}

We denote the fixed transmit power of the MD by $P_{MD}$. Then, the uplink data rate for offloading the data $O_{k,i}, (k,i)\in\mathcal{E},$ to the AP is
\begin{align}\label{rate_up}
R_{k,i}^{u}=W\log_{2}\left(1+\frac{P_{MD}h_{k,i}^u}{\sigma^{2}}\right),
\end{align}
and the corresponding uplink transmission time is
\begin{align}\label{t_trans}
\tau_{k,i}^{u}=\frac{O_{k,i}}{R_{k,i}^{u}}.
\end{align}
The transmission energy consumption is
\begin{align}
e_{k,i}^{u}=\tau_{k,i}^{u}P_{MD}.
\end{align}

We assume that the edge server has $\rho^{c}$ cores and can compute $\rho^{c}$ tasks in parallel. The execution time of task $i$ on the AP is given by
\begin{align}
\tau_{i}^{c}=\frac{L_{i}}{f^{c}},
\end{align}
where $f^{c}$ is the fixed service rate of each CPU core.
Similarly, we can calculate the ready time of task $i$ executed at the edge server as
\begin{align}\label{RT_edge}
RT_{i}^{c}=\max_{k\in\textbf{pred(i)}}\left\{(1-a_{k})\left(FT_{k}^{l}+\tau_{k,i}^{u}\right)+a_{k}FT_{k}^{c}\right\},
\end{align}
and its finish time is
\begin{align}\label{FT_edge}
FT_{i}^{c}=RT_{i}^{c}+\tau_{i}^{c}.
\end{align}

\subsection{Problem Formulation}

We assume that both the MD and MEC server have a lot more CPU cores than needed to execute the possibly concurrent tasks in the considered mobile application. As such, we can safely set $\rho^{l}=\rho^{c}=\infty$. {\color{blue}{Besides, it is assumed that the number of available channels is sufficiently large to execute the possibly concurrent data transmissions in the task graph.}}




From the above discussion, the total time to complete the all tasks is equal to the local finish time of the auxiliary exit task $M+1$, i.e., $FT_{M+1}^{l}$. Besides, we can calculate the total energy consumption of the MD by
\begin{align}\label{energy}
E=\sum_{i=1}^{M}(1-a_{i})e_{i}^{l}+\sum_{i=1}^{M}\sum_{k\in\textbf{pred(i)}}(1-a_{k})a_{i}e_{k,i}^{u},
\end{align}
which consists of energy consumed on local computation and task offloading.

In this paper, we consider the energy-time cost (ETC) as the performance metric, which is defined as the weighted sum of the total energy consumption and execution time, i.e.,
\begin{align}
\eta=\beta_{e}E+\beta_{t}FT_{M+1}^{l},
\end{align}
where $0<\beta_{e}<1$ and $0<\beta_{t}<1$ denote the weights of energy consumption and computation completion time of the MD, respectively. It is assumed that the weights are related by $\beta_{t}=1-\beta_{e}$. {\color{blue}{We consider the weighted-sum approach [9,17,18] for a general multi-objective optimization problem. According to the Proposition 3.9 of \cite{add4}, for any given positive weights, we can reach an efficient solution of the multi-objective optimization problem by solving Problem (P1). A weakly efficient solution will be obtained if any of the weights is zero. Besides, in order to meet user-specific demands, we allow the MD to choose different weights. For instance, the MD with low battery energy prefers a larger $\beta_e$ for energy saving, while for the delay-sensitive MD, a larger $\beta_t$ will be chosen to reduce the execution time.}}

{\color{blue}{Evidently, a higher CPU frequency leads to shorter task execution time. Meanwhile, according to \eqref{local_energy}, the energy consumption per CPU cycle is a quadratic function of the CPU frequency, thus the energy consumption increases with the CPU frequency for executing a task. Because the AP has stable power supply, it can operate with a fixed maximum frequency $f^c$ to minimize the execution delay. However, since the MD is often energy-constrained, we can apply dynamic voltage and frequency scaling (DVFS) technique to tune the local CPU frequency for balancing the performance between energy consumption and execution time.}} Denoting $\mathbf{a}\triangleq\{a_{i}\}$ and $\mathbf{f}\triangleq\{f_{i}^{l}\}$, $i\in\mathcal{M}$, we aim to minimize the ETC of the MD subject to the peak CPU frequency constraint of the MD, i.e.,
\begin{eqnarray}
(P1)~~~~\min_{(\mathbf{a},\mathbf{f})}&&\eta,\nonumber \\
{\rm s.t.}&&0\leq f_{i}^{l}\leq f_{peak},\nonumber \\
&&a_{i}\in\{0,1\}, \forall i\in\mathcal{M},
\end{eqnarray}
where we assume $f^c>f_{peak}$ in this paper.  In general, $(P1)$ is non-convex due to the binary variables $\mathbf{a}$ and the recursive structure of $FT_{M+1}^{l}$. In the following section, we first simplify $(P1)$ by exploiting the property of the total task completion time $FT_{M+1}^{l}$. Then, we propose an efficient method to obtain the optimal CPU frequencies with a given $\mathbf{a}$.

\section{Optimal Resource Allocation Under Fixed Offloading Decisions}

\subsection{Problem (P1) Simplification}

We denote a path $o$ as an ordered sequence of task indices $\Psi(o)=\{k_0^{o},k_{1}^{o},...,k_{m}^{o},...,k_{m_{o}}^{o},k_{m_{o}+1}^{o}\},$ $k_0^{o}=0, k_{m_{o}+1}^{o}=M+1$, that pass through the general task graph $G$ from the entry task $0$ to the exit task $M+1$. Here, $m_o$ is the total number of real tasks in path $o$. For instance, $\{0,1,5,8,10\}$ is a path in Fig. 2. There are three real tasks $\{1, 5, 8\}$ in the path. Besides, we denote the set of all loop-free paths as $\mathcal{O}$, which can be obtained by running the $K$-shortest path routing algorithm on $G$. Likewise, we denote by $O = |\mathcal{O}|$ the total number of paths. Let $T_{o}$ denote the total execution time in the $o$-th path excluding the waiting time for the data inputs from the other paths.  Then, we have
\begin{align}\label{T_o}
T_{o}=\sum_{k^{o}_m\in\Psi(o)}[(1-a_{k^{o}_{m}})\tau_{k^{o}_{m}}^{l}+a_{k^{o}_{m}}\tau_{k^{o}_{m}}^{c}]+\sum_{k^{o}_m=k^{o}_{1}}^{k^{o}_{m_{o}+1}}a_{k^{o}_{m}}(1-a_{k^{o}_{m-1}})\tau_{k^{o}_{m-1},k^{o}_{m}}^{u}+(1-a_{k^{o}_{m}})a_{k^{o}_{m-1}}\tau_{k^{o}_{m-1},k^{o}_{m}}^{d},
\end{align}
which consists of the total computation and communication delay in path $o$.

To simplify Problem (P1), we first have the following lemma on $FT_{M+1}^{l}$.

\emph{\textbf{Lemma 3.1:}} $FT_{M+1}^{l}=\max\{T_1,T_2,...,T_o,...,T_O\}$ holds given any $(\mathbf{a},\mathbf{f})$.

\begin{proof}
Please refer to Appendix \ref{appendicesC}.
\end{proof}

Lemma 3.1 indicates that the final completion time is equal to the largest total execution time of all the paths in $G$. Note that although $T_o$ does not include the time spent on waiting for the task input data from other paths, the largest $T_o$ among all paths is the final completion time.

Due to the one-to-one mapping between $f_{i}^{l}$ and $\tau^l_{i}$ in \eqref{f_mapping}, it is equivalent to optimize (P1) over the time allocation $\tau^l_{i}$. By introducing an auxiliary variable $T_{max}=\max\{T_{1},T_{2},...,T_{o},...,T_{O}\}$, (P1) can be equivalently expressed as
\begin{eqnarray}
(P2)~~~~\min_{(\mathbf{a},\{\tau^l_{i}\},T_{max})}&&\beta_{e}E+\beta_{t}T_{max},\nonumber \\
{\rm s.t.}&&T_{max}\geq T_{1}, T_{max}\geq T_{2},..., T_{max}\geq T_{O},\nonumber \\
&&0\leq \frac{L_i}{\tau_{i}^l}\leq f_{peak},\nonumber \\
&&a_{i}\in\{0,1\}, \forall i\in\mathcal{M}.
\end{eqnarray}

Notice that (P2) is non-convex in general due to the binary variables $\mathbf{a}$. However, for any given $\mathbf{a}$, the remaining optimization over $\{\tau^l_{i}\}$ is a convex problem. In the following, we assume a fixed offloading decision $\mathbf{a}$ and derive an efficient algorithm to obtain the optimal $(\tau^l_{i})^*$, or equivalently the optimal local CPU frequencies $(f_i^l)^*$.

\subsection{Optimal Local CPU Frequencies}

Suppose that $\mathbf{a}$ is given. We express a partial Lagrangian of Problem (P2) as
\begin{align}\label{lag}
L(\{\tau^l_{i}\},T_{max},\lambda_1,...,\lambda_O)=\beta_{e}E+\beta_{t}T_{max}+\sum_{o=1}^O\lambda_o(T_{o}-T_{max}),
\end{align}
where $\{\lambda_o\geq 0,o\in\mathcal{O}\}$ denotes the dual variables associated with the corresponding constraints. Let $\{\lambda_o^{*},o\in\mathcal{O}\}$ denote the optimal dual variables. Then, we derive the closed-form expressions for the optimal local CPU frequencies as follows.

\emph{\textbf{Proposition 3.1:}} $\forall i$ with $a_i=0$, by denoting the index set of the paths that contain task $i$ as $\Upsilon(i)$, the optimal CPU frequencies at the MD satisfy
\begin{align}\label{optimal_f}
(f_{i}^{l})^{*}=\min\left\{\sqrt[3]{\frac{\sum_{o\in\Upsilon(i)}\lambda_{o}^{*}}{2\kappa\beta_{e}}},f_{peak}\right\}.
\end{align}

\begin{proof}
Please refer to Appendix \ref{appendicesA}.
\end{proof}

From Proposition 3.1, we observe that the optimal $(f_{i}^{l})^{*}$ is determined by the dual variables $\lambda_o^*$ corresponding to all the paths containing task $i$. Besides, increasing $\beta_e$ leads to a lower optimal $(f_{i}^{l})^{*}$ for energy saving.

\emph{\textbf{Corollary 3.1:}} The summation of the optimal dual variables over all paths is equal to the constant $\beta_t$. That is,
\begin{align}\label{dual}
\sum_{o\in\mathcal{O}}\lambda_{o}^{*}=\beta_{t}.
\end{align}
Then, if $\Upsilon(i)=\mathcal{O}$, according to the Proposition 3.1, the optimal local CPU frequency for task $i$ is
\begin{align}\label{f_constant}
(f_{i}^{l})^{*}=\min\left\{\sqrt[3]{\frac{\beta_t}{2\kappa\beta_{e}}},f_{peak}\right\},
\end{align}
which is a constant regardless of the values of  $\lambda_o^{*},o\in\mathcal{O}$.

{\color{blue}{
\begin{proof}
Please refer to Appendix \ref{addappendix}.
\end{proof}
}}

The above corollary indicates that the optimal $(f_{i}^{l})^{*}$ is a constant when the $i$-th task is included in all the paths, i.e., $\Upsilon(i)=\mathcal{O}$.

Based on Proposition 3.1 and Corollary 3.1, we can apply the projected subgradient method \cite{convex} to search for the optimal dual variables $\{\lambda_o^{*},o\in\mathcal{O}\}$. Specifically, we initialize $\{\lambda_o^{(0)}\geq 0,o\in\mathcal{O}\}$ satisfying \eqref{dual}. In the $\psi$-th iteration, we first calculate $T_o,\forall o\in\mathcal{O},$ using \eqref{T_o} and \eqref{optimal_f} and set $T_{max}=\max\{T_1,...,T_O\}$. Then, the dual variables are updated to $\{\hat{\lambda}_o^{(\psi)},o\in\mathcal{O}\}$ by using subgradients $(T_o-T_{max}),\forall o\in\mathcal{O}$, i.e.,
\begin{align}\label{subg}
\hat{\lambda}_o^{(\psi)}=\lambda_o^{(\psi-1)}-\epsilon(T_o-T_{max}),
\end{align}
where $\epsilon$ is a small learning rate.
In order to guarantee the feasibility of dual variables, we need to project $\{\hat{\lambda}_o^{(\psi)},o\in\mathcal{O}\}$ to the feasible region given in \eqref{dual}. The projection is calculated from the following convex problem,
\begin{eqnarray}
\min_{\{\lambda_o^{(\psi)}\}}&&\sqrt{\sum_{o\in\mathcal{O}}(\lambda_o^{(\psi)}-\hat{\lambda}_o^{(\psi)})^2},\nonumber \\
{\rm s.t.}&&\sum_{o\in\mathcal{O}}\lambda_{o}^{(\psi)}=\beta_{t},\nonumber\\
&&\lambda_o^{(\psi)}\geq 0, \forall o\in \mathcal{O},\label{project}
\end{eqnarray}
which can be efficiently solved by general convex optimization techniques, e.g.,
interior point method \cite{convex}.
After updating the dual variables, we can further obtain the updated optimal local CPU frequencies. Such iteration proceeds until a stopping criterion is met.
The pseudo-code of the method is shown in Algorithm 1.

\begin{algorithm}[h]
\caption{Optimal algorithm for (P2) under fixed offloading decision }
\begin{algorithmic}[1]
\STATE \textbf{initialize} $\{\lambda_o^{(0)}\geq 0\}$ satisfying \eqref{dual}.
\REPEAT
\STATE Compute $T_o,\forall o\in\mathcal{O},$ using \eqref{T_o} and \eqref{optimal_f} with given $\{\lambda_o^{(\psi)}\}$.
\STATE Set $T_{max}=\max\{T_1,...,T_O\}$.
\STATE Update $\{\lambda_o^{(\psi)}\}$ to $\{\hat{\lambda}_o^{(\psi+1)}\}$ using \eqref{subg}.
\STATE Project $\{\hat{\lambda}_o^{(\psi+1)}\}$ to the feasible region by solving Probelm \eqref{project}.
\UNTIL{$\{\lambda_o\}$ converge to a prescribed accuracy.}
\STATE Obtain $\{(f_{i}^{l})^*\}$ by \eqref{optimal_f}.
\end{algorithmic}
\end{algorithm}

\section{Deep Reinforcement Learning Based Task Offloading}

In the last section, we efficiently obtain the optimal $\mathbf{f}$ given the offloading decision $\mathbf{a}$. Intuitively, we can enumerate all $2^M$ feasible $\mathbf{a}$ and choose the optimal one that achieves the minimum objective of (P2). However, such brute-force search is computationally prohibitive, especially when the problem needs to be frequently re-solved with time-varying channel gains and available server computing power.  Besides, other searching based methods, such as branch-and-bound and Gibbs sampling algorithms, are also time consuming when $M$ is large.

In this section, we propose a DRL-based algorithm to solve the joint optimization under time-varying channel gains and CPU frequency at the edge server. Our goal is to derive an offloading decision policy $\pi$ that can quickly predict an optimal offloading action $\mathbf{a}^*\in\{0,1\}^{M}$ of (P2) once the channel gain $\mathbf{h}=\{h_{k,i}^u, h_{k,i}^d, (k,i)\in\mathcal{E}\}$ and the CPU frequency $f^c$ at the edge server are revealed at the beginning of the execution of the application (task graph). The offloading decision policy is denoted as
\begin{align}
\pi:\{\mathbf{h}, f^c\}\mapsto \mathbf{a}^*.
\end{align}

The algorithm structure is illustrated in Fig. 3. There are two stages in the DRL-based offloading algorithm: one is referred to as the actor-critic network based offloading action generation, and the other is offloading policy update, which are detailed as follows. Furthermore, we propose the one-climb policy to speed up the learning process.

\begin{figure}
\begin{centering}
\includegraphics[scale=0.56]{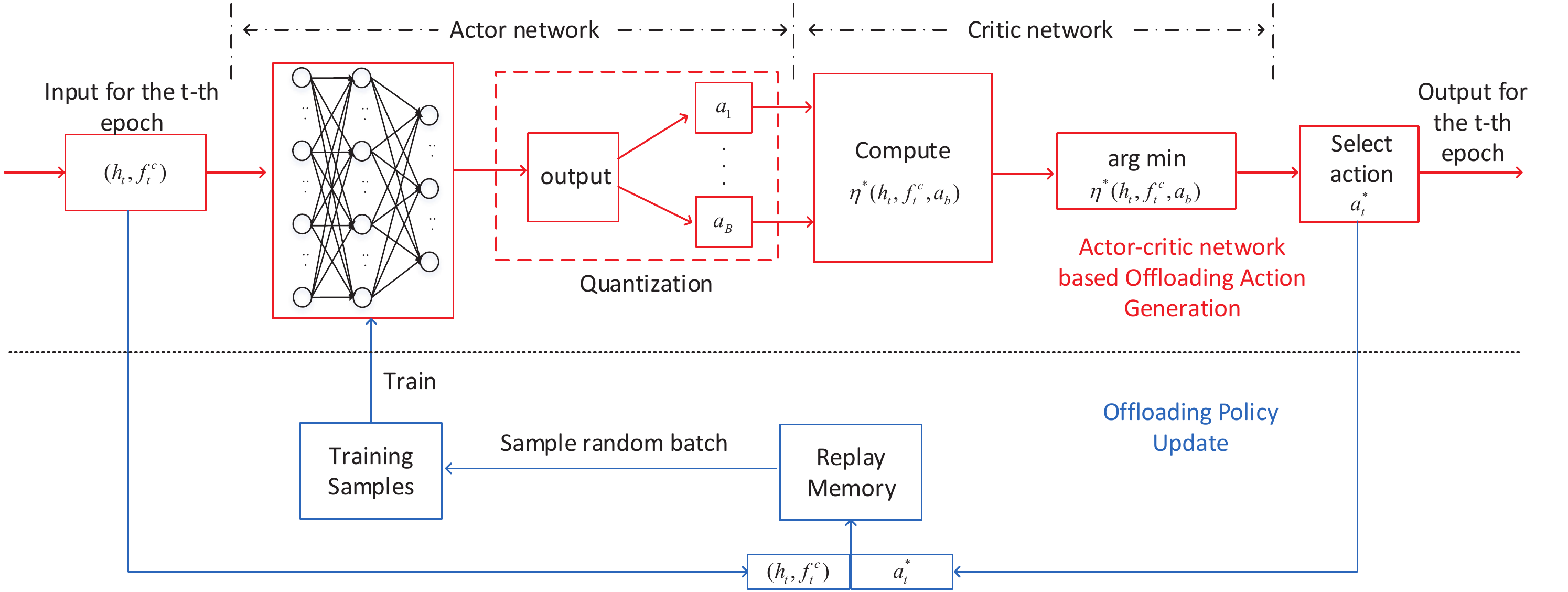}
\vspace{-0.1cm}
 \caption{ The schematics of the deep reinforcement learning framework. }
\end{centering}
\vspace{-0.1cm}
\end{figure}

\subsection{Actor-critic Network Based Offloading Action Generation}

\subsubsection{Actor Network}
The offloading action is generated based on a DNN. We denote the embedded parameters of the DNN at the $t$-th epoch as $\theta_t, t=1,2,...$, where $\theta_1$ is randomly initialized  following a zero-mean normal distribution. At the $t$-th epoch, we take the channel gain $\mathbf{h}_t$ and edge CPU frequency $f^{c}_t$ as the input of the DNN. Accordingly, the DNN outputs a relaxed offloading action $\bar{\mathbf{a}}_t$, which is denoted by a mapping $g_{\theta_t}$, i.e.,
\begin{align}
\bar{\mathbf{a}}_{t}=g_{\theta_{t}}(\mathbf{h}_{t}, f^c_{t}),
\end{align}
where $\bar{\mathbf{a}}_{t}=\{\bar{a}_{t,i}\in[0,1], i=1,...,M\}$, and the $\bar{a}_{t,i}$ denotes the $i$-th entry of $\bar{\mathbf{a}}_{t}$.

%

Notice that each entry of $\bar{\mathbf{a}}_t$ is a continuous value between 0 and 1. To generate a feasible binary offloading decision, we first  quantize $\bar{\mathbf{a}}_{t}$ into $B$ candidate binary offloading actions. Then, the critic network will evaluate the performance of the $B$ candidate actions, and the one with the lowest ETC will be selected as the output solution. Noticeably, for a good quantization method, we only need to generate few candidate actions to reduce the computational complexity. Meanwhile, the quantized actions based on the relaxed action should contain sufficient diversity to yield a lower ETC.  In this paper, we propose a Gaussian noise-added order-preserving (GNOP) quantization method as shown in Fig. 4. We define the quantization function as
\begin{align}
G_{B}:\bar{\mathbf{a}}\mapsto\Omega_t=\{\mathbf{a}_b|\mathbf{a}_b\in\{0,1\}^{M}, b=1,...,B\},
\end{align}
where $\Omega_t$ is the generated candidate action set in the $t$-th epoch.
\begin{figure}
\begin{centering}
\includegraphics[scale=0.9]{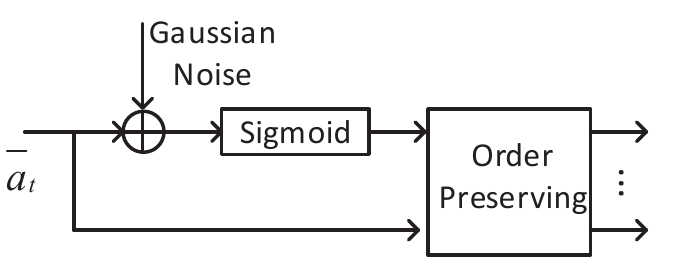}
\vspace{-0.1cm}
 \caption{ GNOP quantization method. }
\end{centering}
\vspace{-0.1cm}
\end{figure}

Order-preserving quantization method was originally introduced to explore the output of the DNN in \cite{DRL4}. The key idea is to preserve the ordering of all the entries in a vector before and after quantization. In our proposed GNOP method, the first $B/2$ actions is generated by traditional order-preserving method, where we assume that $B$ is an even number without loss of generality. Specifically, suppose that the output offloading action is $\bar{\mathbf{a}}_{t}$. The generation rule for $\{\mathbf{a}_{b}, b=1,..., B/2\}$ in the order-preserving method is shown as follow.

First, we obtain the offloading decision $\mathbf{a}_{1}$ as
\begin{align}
a_{1,i}=\left\{
                 \begin{array}{ll}
                   1, & \bar{a}_{t,i}>0.5, \\
                   0, & \bar{a}_{t,i}\leq0.5,
                 \end{array}
               \right.
\end{align}
for $i=1,...,M$. For the other $B/2-1$ offloading actions, we first order the entries of $\bar{\mathbf{a}}_{t}$ according to their distances to 0.5, i.e., $|\bar{a}_{t,(1)}-0.5|\leq|\bar{a}_{t,(2)}-0.5|\leq...\leq|\bar{a}_{t,(i)}-0.5|\leq...\leq|\bar{a}_{t,(M)}-0.5|$, where $\bar{a}_{t,(i)}$ is denoted as the $i$-th order entry of $\bar{\mathbf{a}}_{t}$. Then, the $b$-th offloading action $\mathbf{a}_{b}$ is obtained as
\begin{align}
a_{b,i}=\left\{
                 \begin{array}{ll}
                   1, & \bar{a}_{t,i}>\bar{a}_{t,(b-1)}, \\
                   1, & \bar{a}_{t,i}=\bar{a}_{t,(b-1)}~\hbox{and}~\bar{a}_{t,(b-1)}<0.5 ,\\
                   0, & \bar{a}_{t,i}=\bar{a}_{t,(b-1)}~\hbox{and}~\bar{a}_{t,(b-1)}>0.5 ,\\
                   0, & \bar{a}_{t,i}<\bar{a}_{t,(b-1)},
                 \end{array}
               \right.
\end{align}
for $i=1,...,M$ and $b=2,...,B/2$.

Compared to the traditional $K$-nearest neighbor (KNN) method, the order-preserving quantization method leads to a higher diversity in the offloading action space. However, the offloading actions produced by conventional order-preserving quantization method are still closely placed around $\bar{\mathbf{a}}_t$, which reduces the chance of finding a local optimum in a large action space. To better explore the action space, we introduce a Gaussian noise-added approach to generate the other half of $B/2$ candidate actions.
Specifically, we first add a Gaussian noise to $\bar{\mathbf{a}}_{t}$ as
\begin{align}
\ddot{\mathbf{a}}_{t}=f_{sg}(\bar{\mathbf{a}}_{t}+\mathbf{n}),
\end{align}
where $\mathbf{n}\thicksim\mathcal{N}(0,1)$ and $f_{sg}(\cdot)$ is the sigmoid function that maps the original noise-added action to $\ddot{\mathbf{a}}_{t}\in[0,1]$. Then, we apply the order-preserving method on $\ddot{\mathbf{a}}_{t}$ to generate the $B/2$ offloading actions.

\subsubsection{Critic Network}
After generating the candidate offloading actions in the actor network, we evaluate the ETC performance of each action in the critic network. Instead of training a critic DNN as the conventional actor-critic method does, we can accurately and efficiently evaluate the ETC corresponding to each candidate $\mathbf{a}_b$ using our analysis in Section III. In particular, we denote the ETC achieved by the candidate $\mathbf{a}_b$ as $\eta^*(\mathbf{h}_t,f^c_t,\mathbf{a}_b)$ by optimizing the local CPU frequencies $\mathbf{f}$ as described in Algorithm 1.  This greatly reduces the training cost of the critic DNN and increases the accuracy of ETC evaluation. Accordingly, we choose the best offloading action $\mathbf{a}_t^*$ at the $t$-th epoch as
\begin{align}
\mathbf{a}_t^*=\arg\min_{\mathbf{a}_b\in\Omega_t}\eta^*(\mathbf{h}_t,f^c_t,\mathbf{a}_b).
\end{align}
Noticeably, $\mathbf{a}_t^*$, together with its corresponding optimal resource allocation $\mathbf{f}^*$ constitutes the optimal solution to Problem (P1) (or equivalently, Problem (P2)).

\subsection{Offloading Policy Update}

The optimal actions learned in the offloading action generation stage are used to update the parameters of the DNN through the offloading policy update stage.

As illustrated in Fig. 3, we implement a replay memory to store the past state-action pairs, where the memory is of limited capacity. At the $t$-th epoch, $(\{\mathbf{h}_{t}, f^c_{t}\},\mathbf{a}^*_t)$ obtained in the actor-critic network based offloading action generation stage is added to the memory as a new training data sample. Note that the newly generated data sample will replace the oldest one if the memory is full.

The data samples stored in the memory are used to train the DNN. Specifically, in the $t$-th epoch, we randomly select a batch of training data samples $\{(\{\mathbf{h}_{\omega}, f^c_{\omega}\},\mathbf{a}^*_\omega),\omega\in\mathcal{T}_t\}$ from the memory, where $\mathcal{T}_t$ represents the set of chosen time indices. Then, we minimize the average cross-entropy loss $Loss(\theta_{t})$ through the Adam algorithm in order to update the parameters $\theta_t$ of the DNN, where
\begin{align}
Loss(\theta_{t})=-\frac{1}{|\mathcal{T}_t|}\sum_{\omega\in\mathcal{T}_t}\left((\mathbf{a}^*_\omega)^{\top}\log g_{\theta_t}(\mathbf{h}_{\omega}, f^c_{\omega})+(1-\mathbf{a}^*_\omega)^{\top}\log (1-g_{\theta_t}(\mathbf{h}_{\omega}, f^c_{\omega}))\right).
\end{align}
$|\mathcal{T}_t|$ is the size of $\mathcal{T}_t$, the superscript $\top$ denotes the transpose operator, and the log function is the element-wise logarithm operation for a vector. For brevity, the detail of the Adam algorithm is omitted here. In practice, we start the training step when the number of samples is larger than half of the memory size and train the DNN in every $\delta$ epochs in order to collect a sufficient number of new data samples in the memory.

\subsection{Low-complexity Action Generation Method}

%
\begin{figure}
\begin{centering}
\includegraphics[scale=0.8]{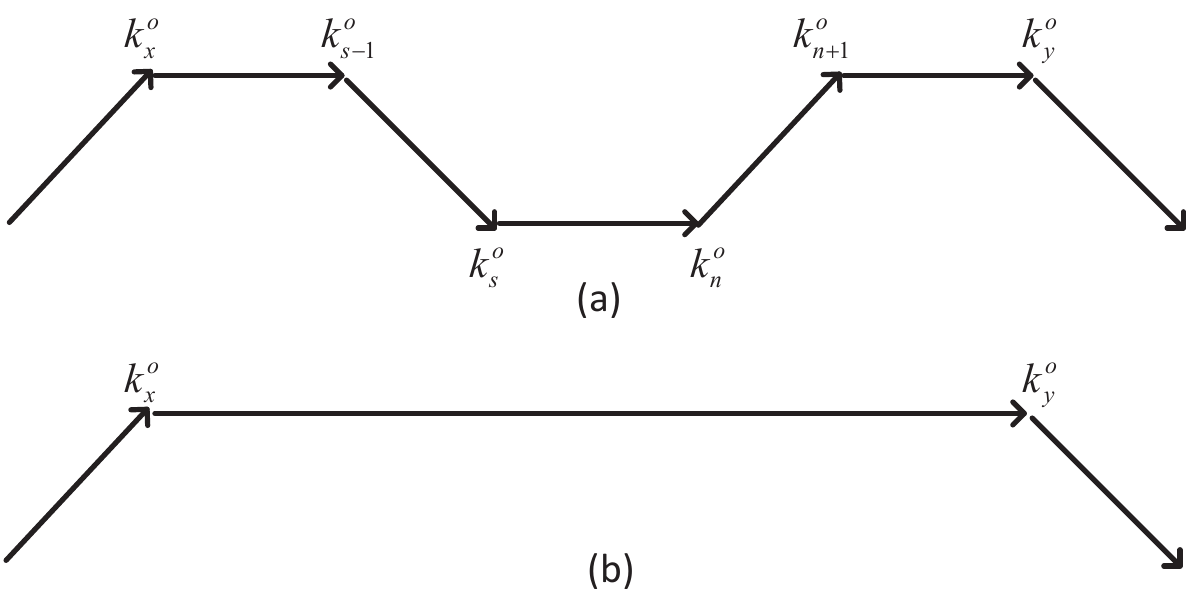}
\vspace{-0.1cm}
 \caption{ Illustration of a two-time offloading and an one-climb schemes in a path $o$. }
\end{centering}
\vspace{-0.1cm}
\end{figure}

Within the proposed DRL framework, we improve the GNOP quantization method to further reduce the complexity.  The basic idea is to restrict our action selection only to those offloading decisions that satisfy the following one-climb policy.

\emph{\textbf{Definition 3 (One-climb policy):}} The execution for the tasks in each path of the graph $G$ migrates at most once from the MD to the edge server.

Fig. 5 illustrates the two-time offloading and one-climb schemes in a path $o$. We show in the Appendix \ref{appendicesB} that by converting the scheme from the two-time offloading to the one-climb policy, the MD saves the energy and time costs for the path $o$. This however may increase the ETC of other paths with overlapping tasks with path $o$. We show that, certain mild conditions hold if the minimum ETC is achieved when all the paths satisfy the one-climb policy. Please refer to Appendix \ref{appendicesB} for the detailed analysis.

The one-climb policy is applied to reduce the number of offloading actions to be evaluated by the critic network. Suppose that $\Omega_t=\{\mathbf{a}_b|\mathbf{a}_b\in\{0,1\}^{M}, b=1,...,B\}$ is the set of actions obtained by the GNOP quantization method at the $t$-th epoch. We remove the actions in $\Omega_t$ that violate the one-climb policy.
By using the one-climb policy in the quantization module, we efficiently reduce the number of calculations for Algorithm 1 at the actor-critic network based offloading action generation stage.


%
%
%

\begin{table}[htb]
\centering
\begin{tabular}{|l|c|r|}
\hline
$W=2\times 10^{6}$ Hz&$\kappa=10^{-26}$\\
\hline
$\sigma^{2}=10^{-10} $ Watt&$f_{peak}=0.01$ GHz\\
\hline
$P_{MD}=0.1 $ Watt&$f^{c}\sim\mathcal{U}(2,50)$ GHz\\
\hline
$P_{AP}=1 $ Watt&$d=20 $ meters\\
\hline
$A_d=4.11$ &$f_c=915 $ MHz\\
\hline
$PL=3$ &$\beta_t=0.5$\\
\hline
$\beta_e=0.5$& \\
\hline
\end{tabular}
\caption{Simulation Parameters}
\end{table}

%

\section{Numerical Results}
In this section, we evaluate the performance of our proposed algorithm through numerical simulations. Consider three different task graphs in Fig. 6, each consisting of 8 actual tasks. Fig. 6(a) illustrates a mesh task graph including a set of linear chains, while a task graph with tree-based structure is considered in Fig. 6(b). In Fig. 6(c), we consider a general task graph which is a combination of the mesh and the tree. The input and output data size (KByte) of each task are shown in Fig. 6. We assume that the computing workload $\{L_i\}=[ 60.5~ 80.3~ 152.6~ 105.8~ 195.3~ 86.4~ 166.8~ 100.3]$ (Mcycles) for all the three task graphs. The transmit power at the MD and the AP are fixed as 100 mW and 1 W, respectively. It is assumed that the CPU frequency $f^c$ is time-varying and follows a uniform distribution between 2 GHz and 50 GHz. Besides, the peak computational frequency of the MD is equal to 0.01 GHz.

In the simulations, we assume that the average channel gain $\bar{h}_{k,i}$ follows the free-space path loss model $\bar{h}_{k,i}=A_{d}(\frac{3\cdot 10^8}{4\pi f_cd})^{PL}$, where $A_d=4.11$ denotes the antenna gain, $f_c=915$ MHz denotes the carrier frequency, $d=20$ in meters denotes the distance between the MD and the AP, and $PL=3$ denotes the pass loss exponent. {\color{blue}{The time-varying fading channel $h_{k,i}^u$ follows an i.i.d. Rician distribution, where the LOS link power is equal to $0.6\bar{h}_{k,i}$. Besides, we follow some classic uplink-downlink channel models that the random variable downlink channel $h_{k,i}^d$ is correlated with the uplink channel $h_{k,i}^u$ and we set the correlation coefficient as 0.7 (the coefficient 0.7  is used in \cite{add5} for modeling weakly-correlated uplink and downlink channels. For some highly correlated case, the correlation coefficient is larger than 0.9).}}
The noise power $\sigma^{2}=10^{-10}$ W. In addition, we set the computing efficiency parameter $\kappa=10^{-26}$, and the bandwidth $W=2$ MHz. The priority weights of energy consumption and computation time of the MD are set as $\beta_{t}=\beta_{e}=0.5$. The parameters used in the simulations are listed in Table I.

We consider a fully connected DNN consisting of one input layer, three hidden layers, and one output layer in the proposed DRL algorithm, where the first, second, and third hidden layers have 160, 120, and 80 hidden neurons, respectively. We implement the DRL algorithm in Python with TensorFlow and set the learning rate for Adam optimizer as 0.01, the training batch size $|\mathcal{T}|=128$, the memory size as 1024, and the training interval $\delta=10$.

\begin{figure*}[!htb]
  \centering
    \subfigure[The mesh task graph.]{\includegraphics[width=0.45\textwidth]{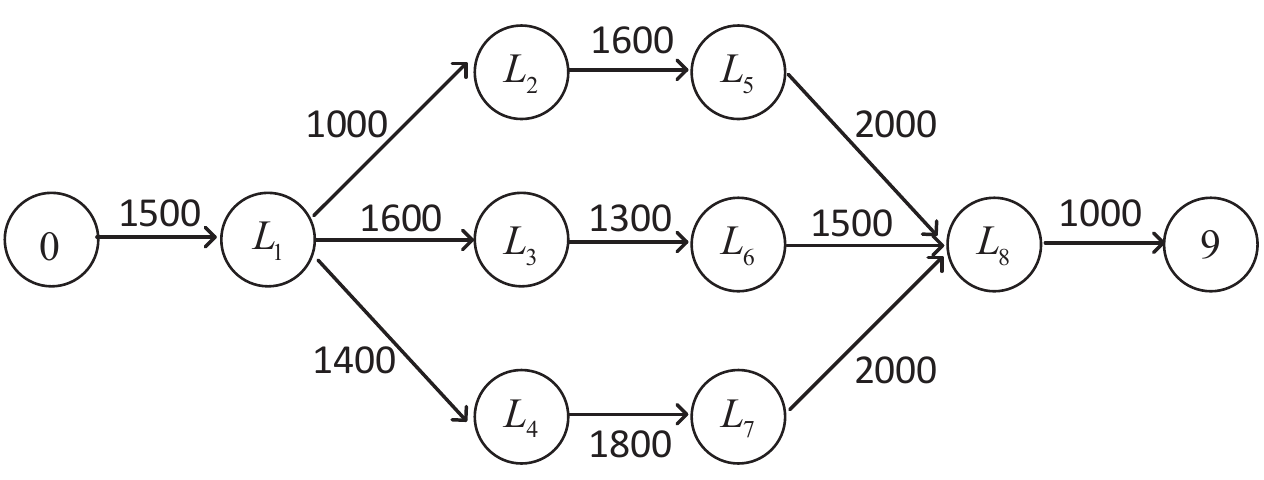}}
    \subfigure[The tree task graph.]{\includegraphics[width=0.45\textwidth]{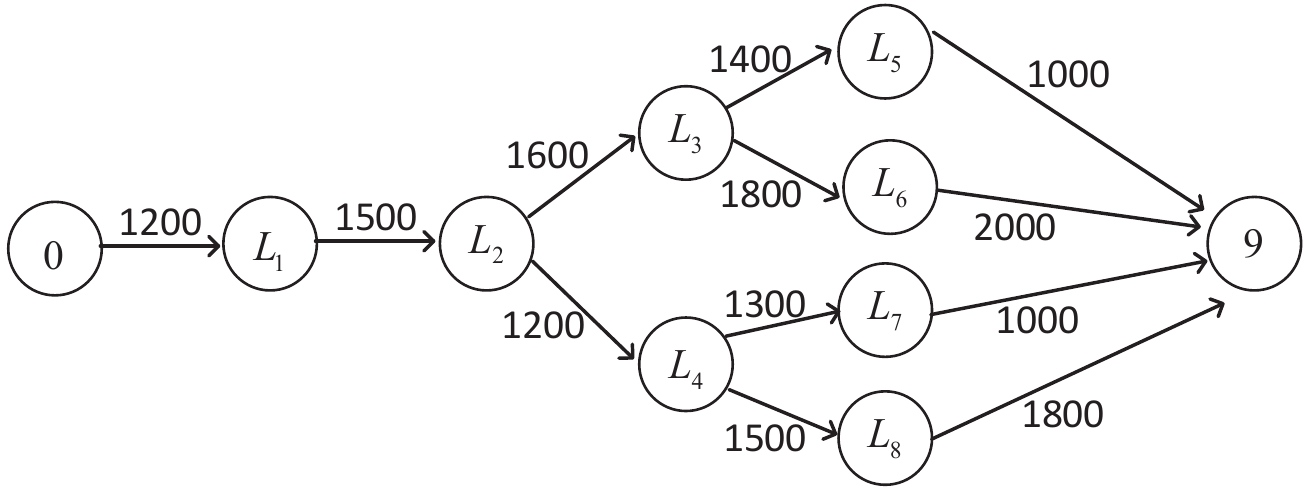}} \\
    \subfigure[The general task graph.]{\includegraphics[width=0.45\textwidth]{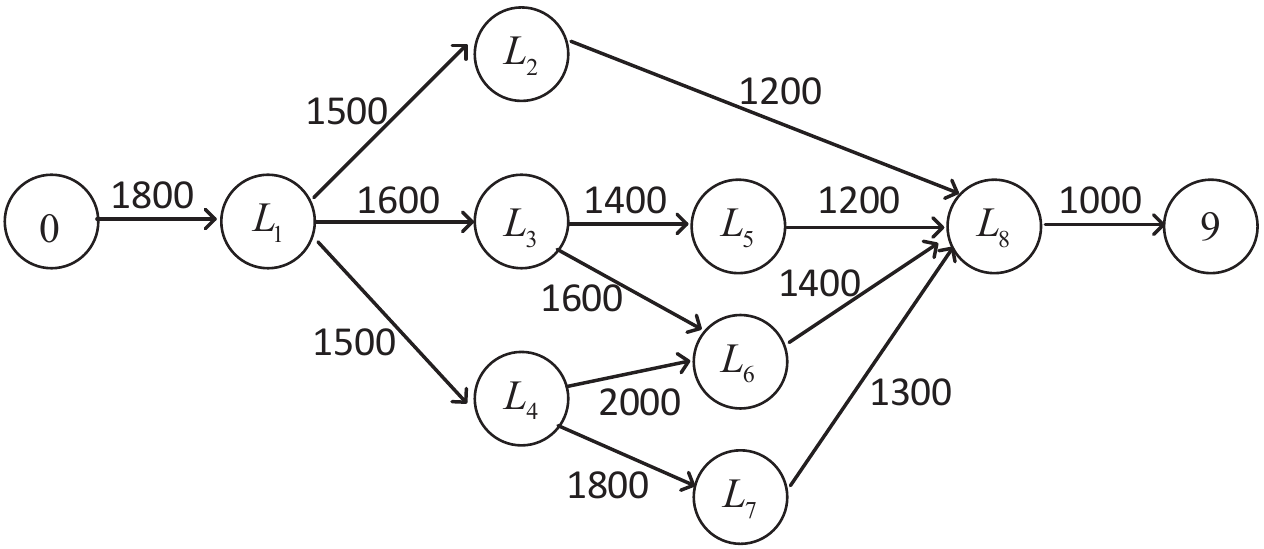}}
     \\
  \caption{The considered task graphs in the simulation.}
\end{figure*}

%
%
\begin{figure*}[t]
  \centering
    \subfigure[With different learning rate.]{\includegraphics[width=0.45\textwidth]{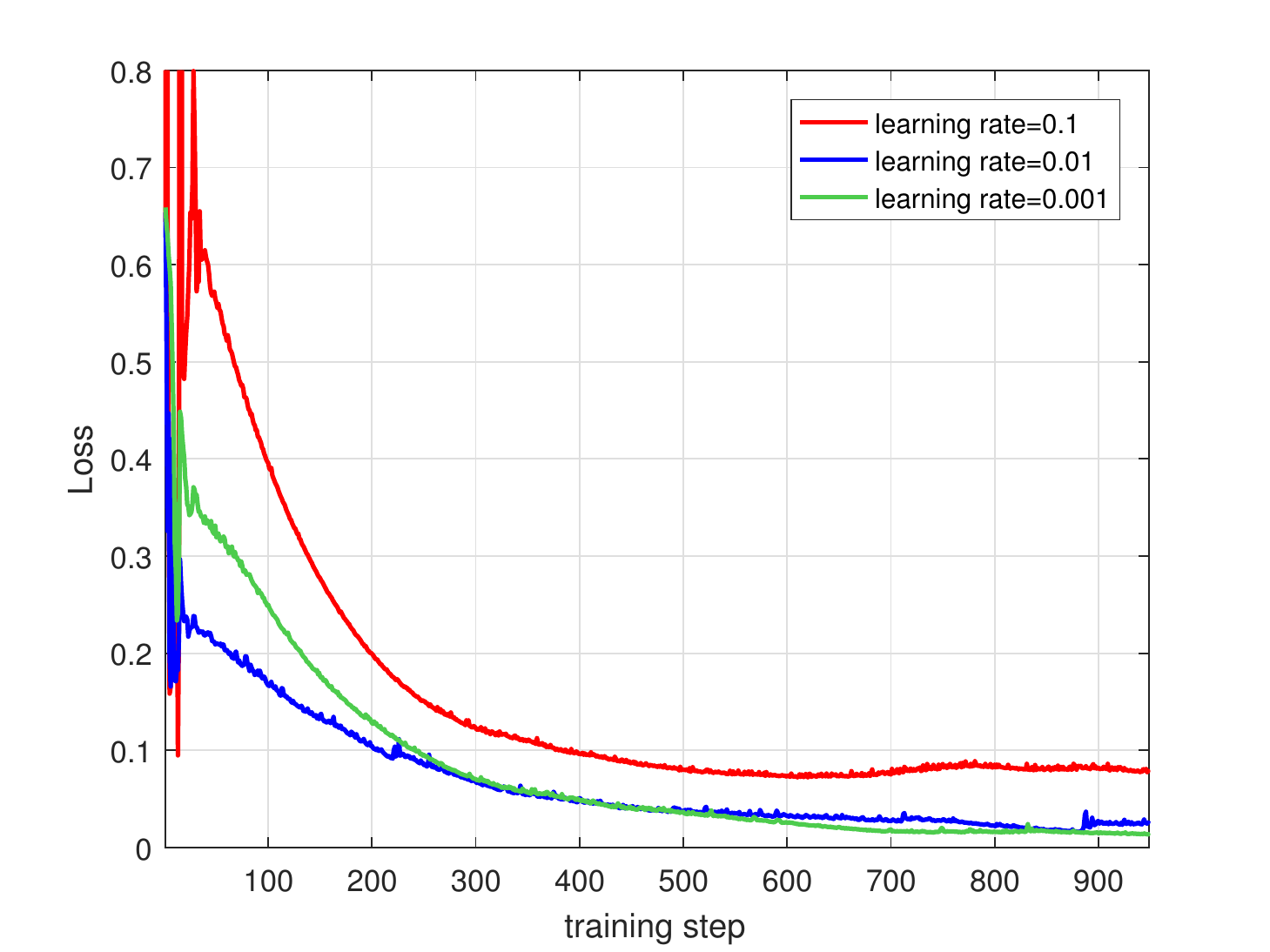}}
    \subfigure[With different batch size.]{\includegraphics[width=0.45\textwidth]{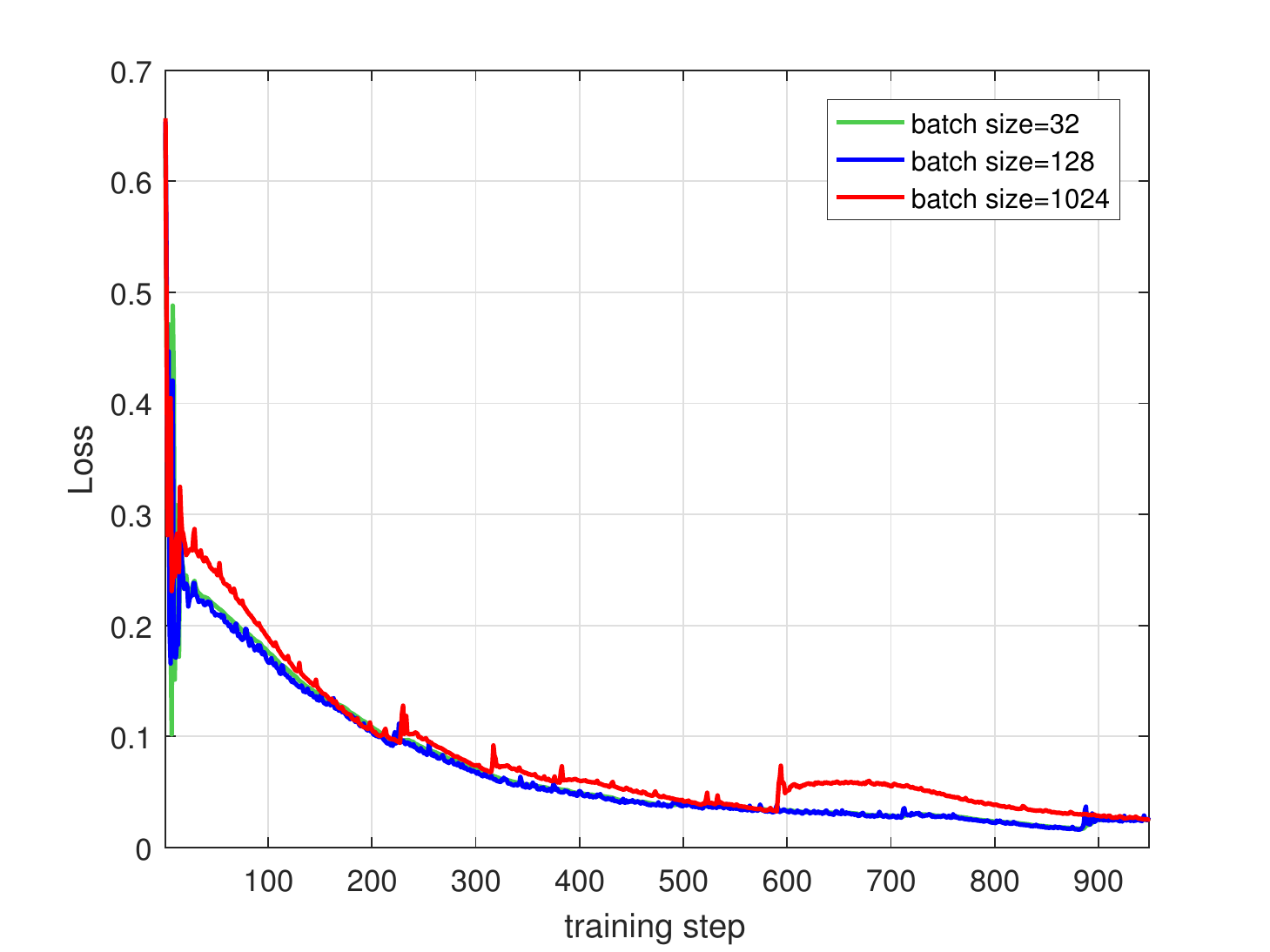}} \\
    \subfigure[With different memory size.]{\includegraphics[width=0.45\textwidth]{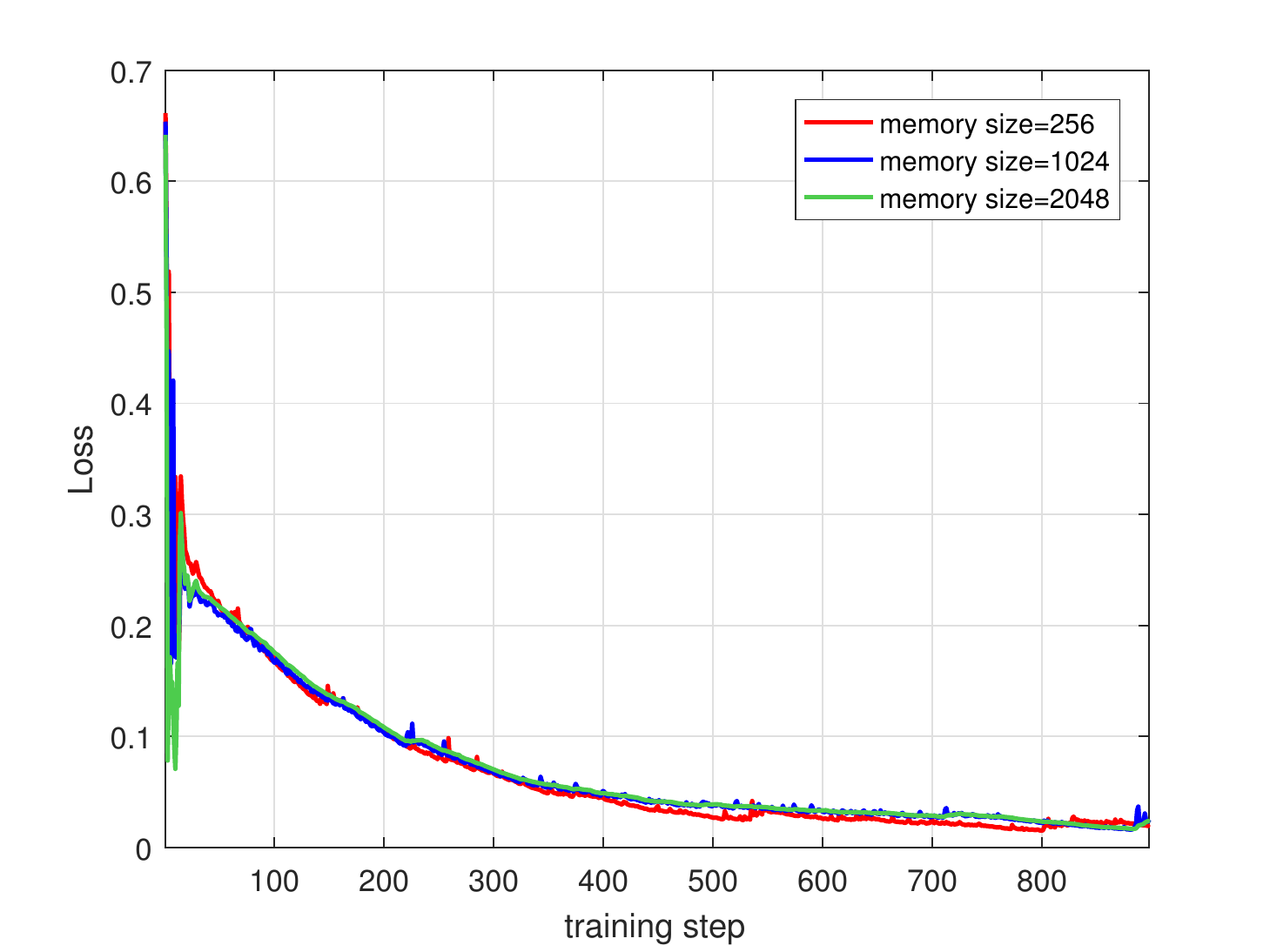}}
    \subfigure[With different training interval.]{\includegraphics[width=0.45\textwidth]{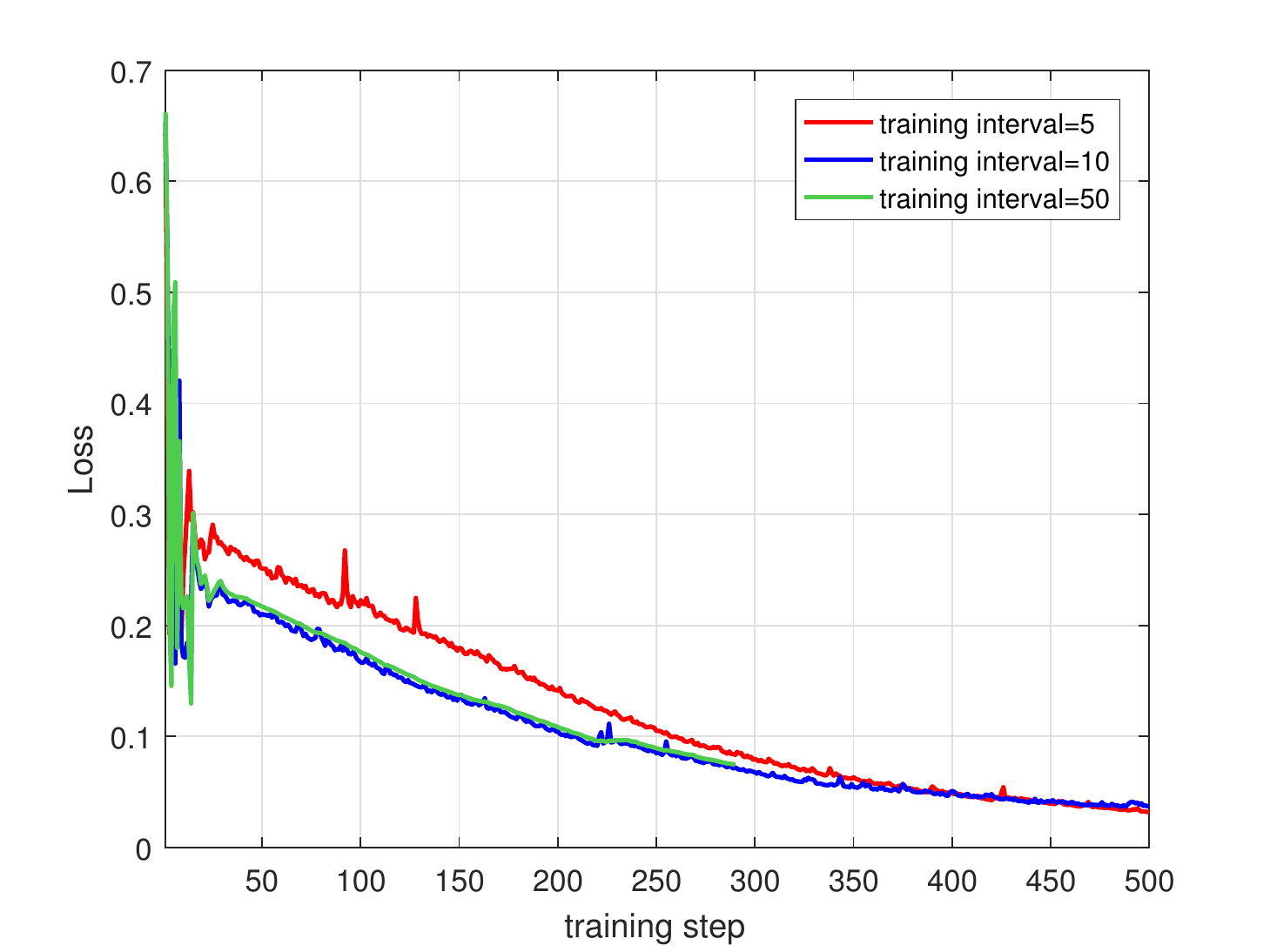}} \\
  \caption{Moving average of the training loss for the tree task graph with different parameters.}
\end{figure*}

\subsection{Convergence Performance}

Without loss of generality, we first consider the tree task graph in Fig. 6(b) as an example to study the impact of the parameters on the convergence performance of the proposed DRL algorithm, including learning rates, batch sizes, memory sizes, and learning intervals in Fig. 7. As shown in Fig. 7(a), we illustrate the impact of the learning rate in Adam optimizer on the moving average of the training loss over moving windows of 15 epochs. It is observed that a too large (i.e., 0.1) or a too small (i.e., 0.001) learning rate leads to a worse convergence. Therefore, in the following simulations, we set the learning rate as 0.01. As for different batch sizes in Fig. 7(b), we observe that a large batch size (i.e., 1024) causes higher fluctuation for the moving average of the training loss, which is due to the frequent usage of the ``old" training data in the memory. Besides, a large batch size consumes more time when training the DNN. Hence, the training batch size is set to 128 in the following simulations. In Fig. 7(c), the moving average of the training loss gradually decreases and stabilizes at around 0.01 for different memory sizes. In addition, we observe that the convergence performance is insensitive to the memory size. In Fig. 7(d), we investigate the convergence of our proposed DRL algorithm under different training intervals. It is observed that for different training intervals, the moving average of the training loss gradually decreases and becomes stable at around 0.02 after 400 training steps, which means that the convergence performance is insensible with respect to the training intervals. In the following simulations, we set the training interval as 10.


%
%
%
Accordingly, Fig. 8 illustrates the convergence performance of the DRL algorithm for the three task graphs, where we set the learning rate as 0.01, the training batch size as 128, the memory size as 1024, and the training interval as 10. We observe that under different task graphs, the moving average of the training loss is below 0.1 after 300 training steps.

{\color{blue}{In Fig. 9, we plot the moving average of the accuracy rates over training steps for the three task graphs, where the proposed DRL algorithm is tested in each training step using 50 independent realizations. We define the accuracy rate as $\chi=1-\frac{\eta_{DRL}-\eta^*}{\eta^*}$, where $\eta^*$ is the average optimal ETC obtained by the exhaustive search method under the 50 independent realizations and  $\frac{\eta_{DRL}-\eta^*}{\eta^*}$ is the ratio of bias of the ETC in DRL algorithm compared to the optimum. We see that the moving average of the accuracy rates for the proposed DRL algorithm gradually converges as the training step increases. Specifically, for the mesh task graph, the achieved $\chi$ exceeds 0.99 after 800 training steps.}}

\begin{figure}
\begin{centering}
\includegraphics[scale=0.6]{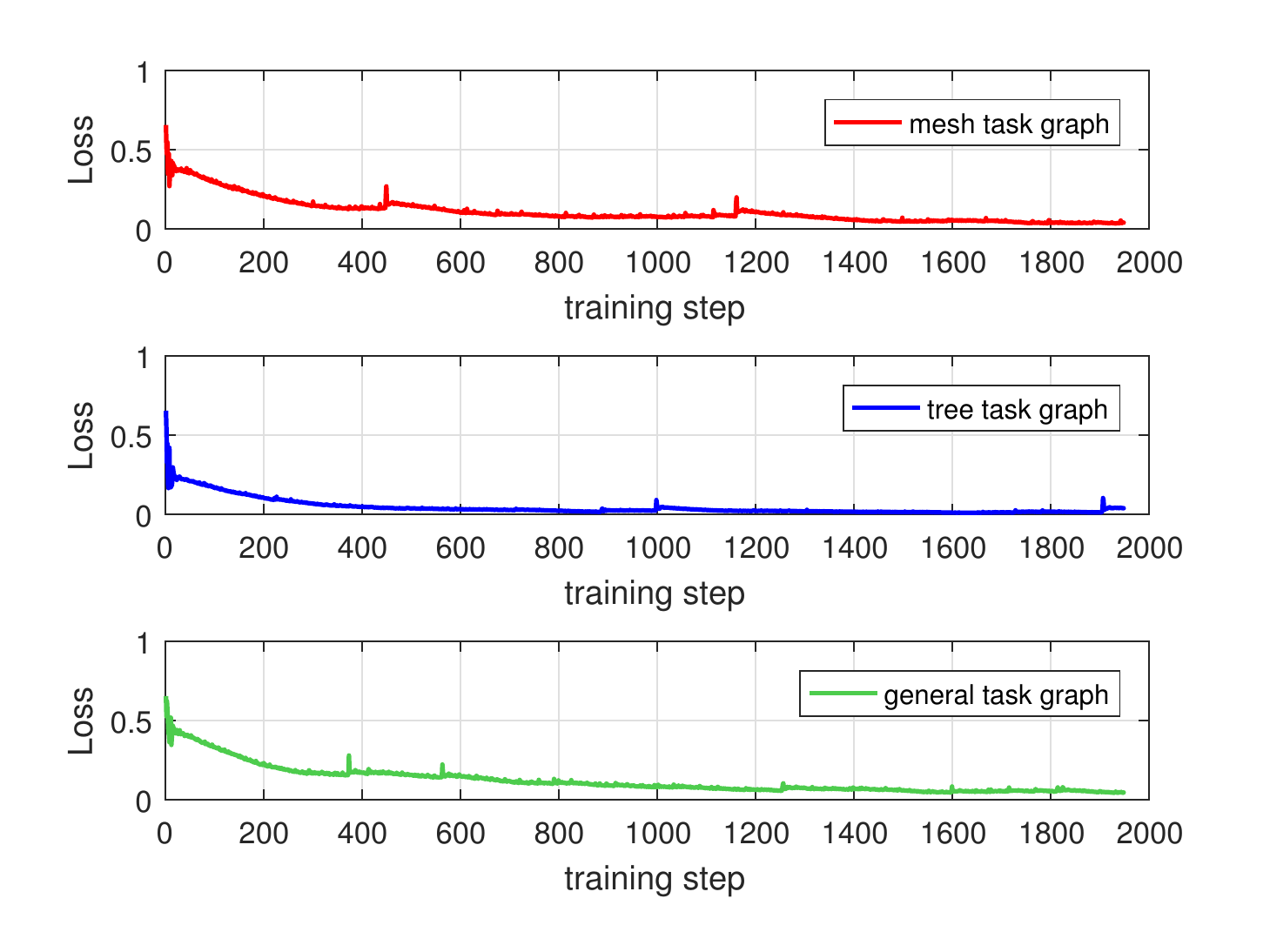}
\vspace{-0.1cm}
 \caption{ Moving average of the training loss for the three task graphs when the learning rate is 0.01, the training batch size is 128, the memory size is 1024, and the training interval is 10. }
\end{centering}
\vspace{-0.1cm}
\end{figure}

\begin{figure}
\begin{centering}
\includegraphics[scale=0.6]{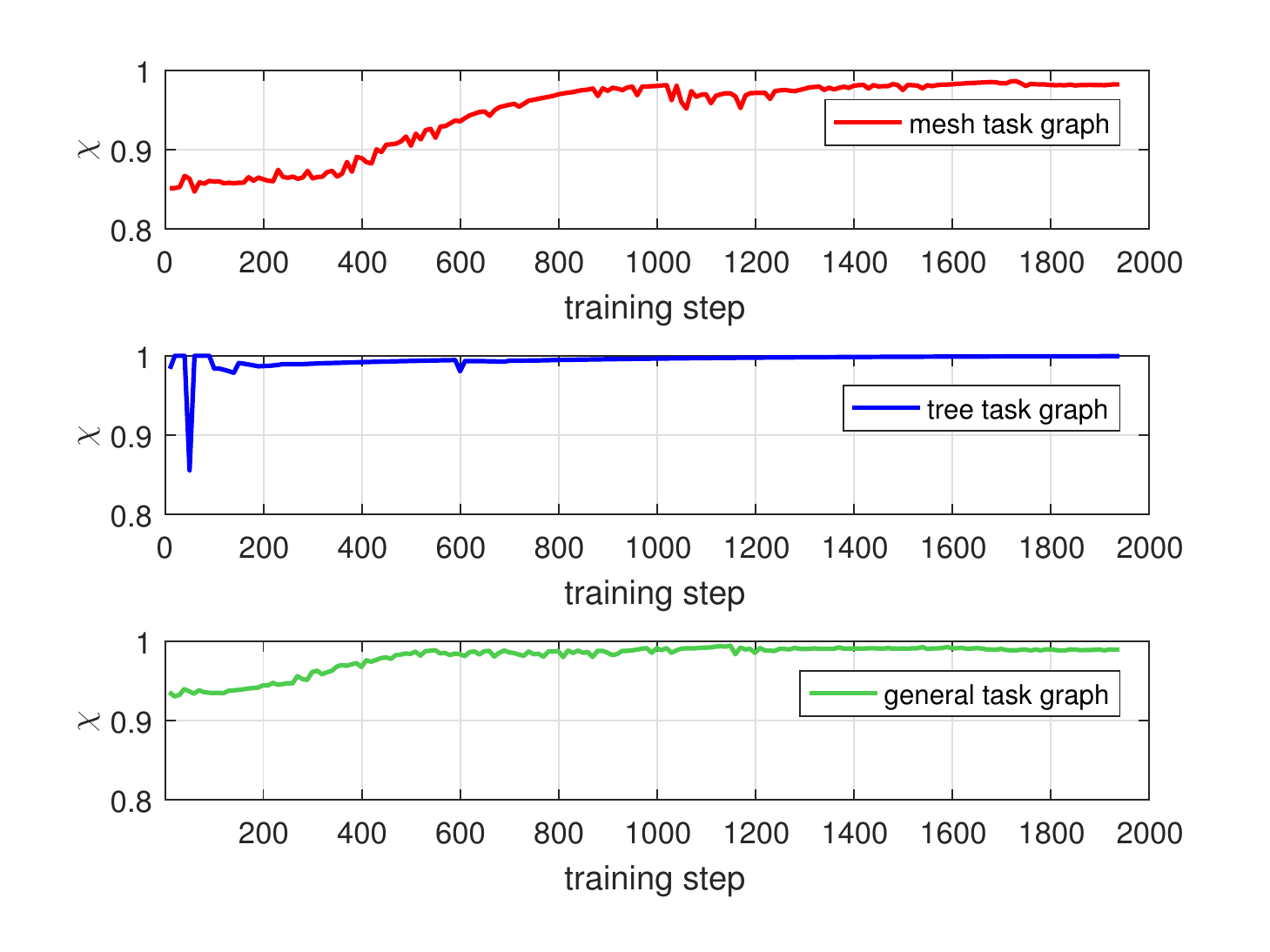}
\vspace{-0.1cm}
 \caption{ Moving average of the accuracy rates over training steps for the three task graphs when the learning rate is 0.01, the training batch size is 128, the memory size is 1024, and the training interval is 10. }
\end{centering}
\vspace{-0.1cm}
\end{figure}

\subsection{Energy and Time Cost (ETC)  Performance Evaluation}

We now compare the energy and time cost (ETC) performance of the proposed methods with that of the following four representative benchmarks.

\begin{itemize}
  \item Gibbs sampling algorithm. The Gibbs sampling algorithm updates the offloading decision iteratively based on the designed probability distribution with respect to the objective values and the temperature parameter. According to the proof in \cite{GS2}, a Gibbs sampling algorithm obtains the optimal solution when it converges.
  \item Exhaustive search. We enumerate all $2^M$ feasible offloading decisions and choose the optimal one that yields the minimum ETC.
  \item All edge computing. In this scheme, all the tasks of the MD are offloaded to the edge side for execution.
  \item All local computing. In this scheme, all the tasks of the MD are executed locally.
\end{itemize}

\begin{figure}
\begin{centering}
\includegraphics[scale=0.6]{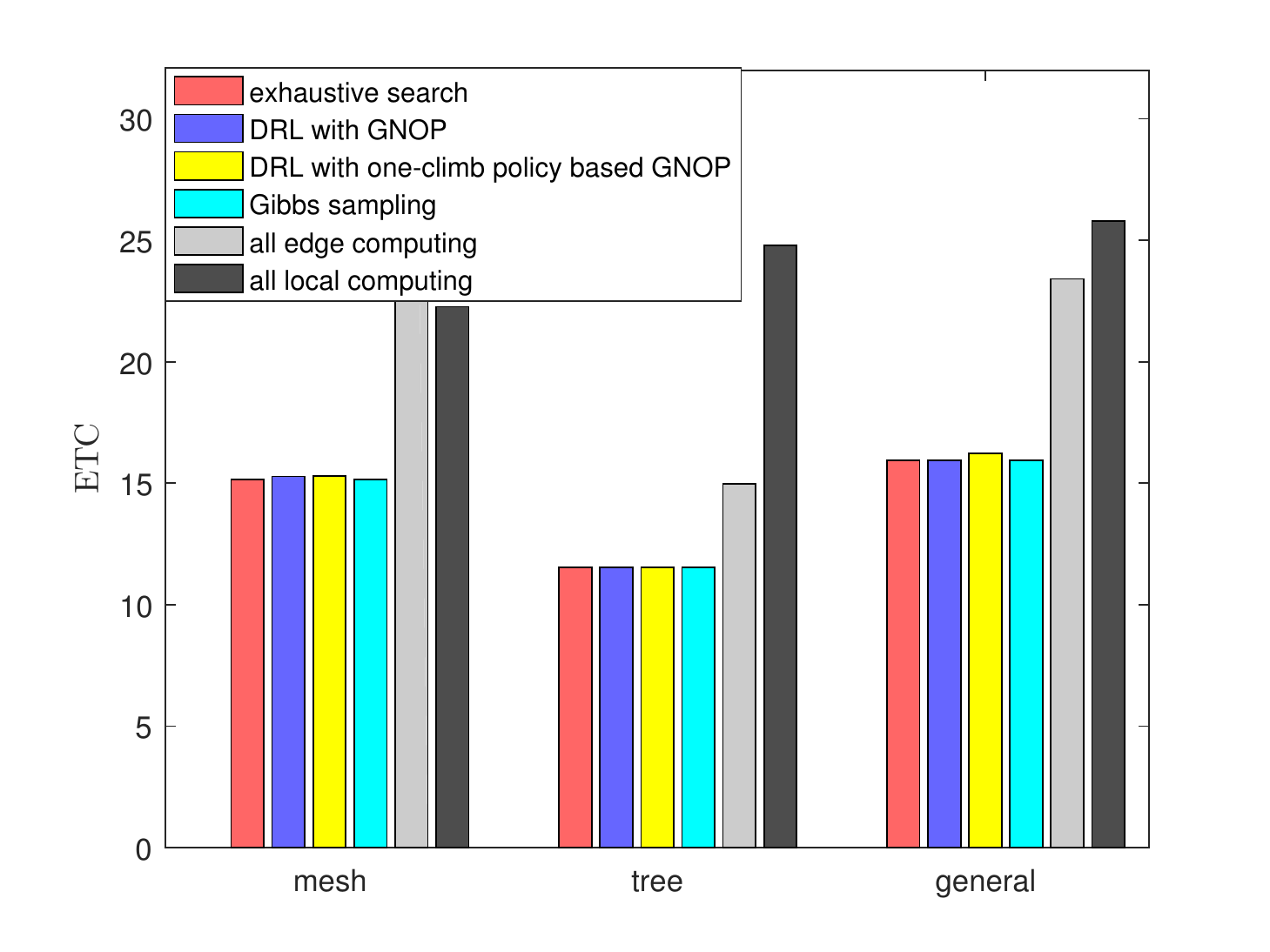}
\vspace{-0.1cm}
 \caption{ Comparisons of ETC performance for different offloading algorithms. }
\end{centering}
\vspace{-0.1cm}
\end{figure}

In Fig. 10, we compare the ETC performance among different offloading schemes under the three task topologies in Fig. 6. Each point in the figure is the average performance of 50 independent realizations. When evaluating the performance, we have neglected the first 20000 time epochs as a warm-up period, so that the DRL has converged. We observe that for all the three task graphs, our proposed DRL algorithm can achieve near-optimal performance compared with the exhaustive search and the Gibbs sampling algorithms. In addition, by applying the one-climb policy heuristics in the GNOP quantization method, the ETC performance is hardly affected. Besides, the DRL algorithm significantly outperforms the all-edge-computing and all-local-computing schemes. This suggests the benefit of adapting the offloading decisions under different wireless channels and edge CPU frequency.

Then, Table II illustrates the average accuracy rates of our proposed DRL algorithm. It is observed that on average the DRL algorithm achieves over $99.1\%$ of the optimal ETC. Specifically, for the general task graph shown in Fig. 6(c), $99.9\%$ accuracy rate with respect to the ETC objective is achieved.


\begin{table}[htb]
\centering
\begin{tabular}{cccc}
\toprule
 & Mesh & Tree & General \\
\midrule
$\chi$& $99.1\%$& $99.9\%$& $99.9\%$ \\
\bottomrule
\end{tabular}
\caption{Accuracy rates $\chi$ for different task graphs.}
\end{table}

\subsection{Complexity of the Proposed DRL Algorithm}

At last, we compare the computational complexity among the four algorithms, where the number of quantized offloading decisions for each epoch in the DRL algorithm $B=16$. We see from the Table III that the DRL algorithm with one-climb policy based GNOP quantization significantly reduces the computation time compared with the DRL algorithm with GNOP method. That is, around $37.15\%$, $4.86\%$, and $33.86\%$ lower average runtime achieved in the mesh, tree, and general task graphs, respectively. Therefore, the one-climb policy heuristics can achieve the near performance as the original GNOP method, while efficiently reducing the complexity of the proposed DRL algorithm.  Specifically, in Fig. 11, we illustrate the computation time for each epoch in the DRL algorithm with one-climb policy based GNOP method under the tree task graph.
For some epochs, the DRL algorithm with one-climb policy based GNOP only consumes around 0.3 second for obtaining the optimal solution.

\begin{figure}
\begin{centering}
\includegraphics[scale=0.6]{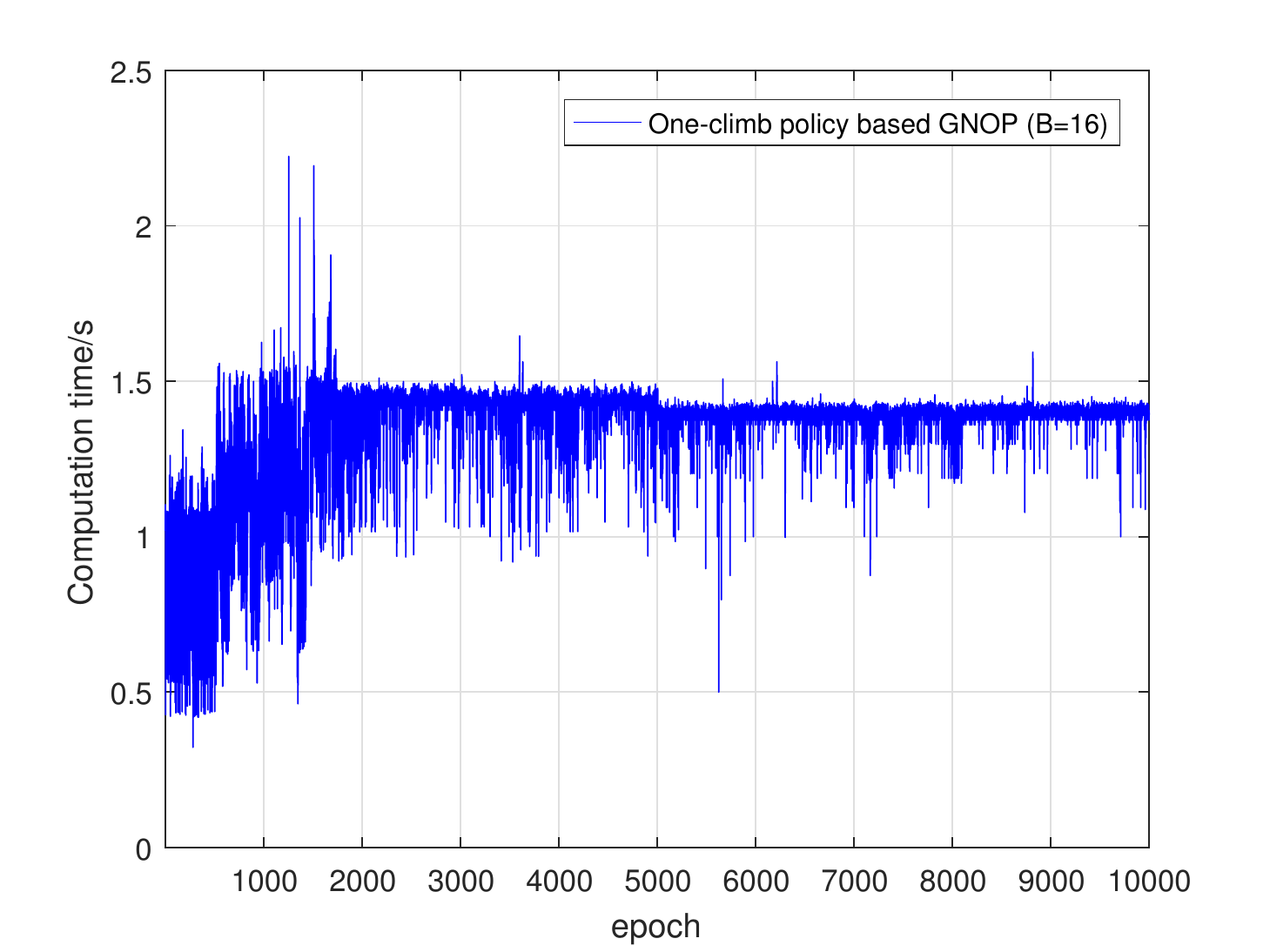}
\vspace{-0.1cm}
 \caption{ Computation time for each epoch under the tree task graph. }
\end{centering}
\vspace{-0.1cm}
\end{figure}

\begin{table}[htb]
\centering
\begin{tabular}{cccc}
\toprule
  &Mesh & Tree & General \\
\midrule
 DRL with One-climb policy based GNOP ($B=16$)&\textbf{0.9240 s} & \textbf{1.3421 s} & \textbf{1.0464 s}  \\
\midrule
DRL with GNOP ($B=16$)&1.4702 s& 1.4107 s & 1.5821 s  \\
\midrule
Gibbs sampling& 8.2039 s  & 8.3046 s & 8.6101 s  \\
\midrule
Exhaustive search& 25.6690 s  & 26.8181 s  & 27.5185 s  \\

\bottomrule
\end{tabular}
\caption{Comparisions of average computation time for each realization.}
\end{table}

Furthermore, as shown in Table III, the DRL algorithm with one-climb policy based GNOP requires much shorter runtime than the Gibbs sampling algorithm and the exhaustive search method. In particular, for the general task graph, it outputs an offloading decision in around 1 second for each realization on average, while the Gibbs sampling and exhaustive search methods spend 8 times and 26 times longer runtime, respectively.

\section{Conclusions}

Considering a single-user MEC system with a general task graph, this paper has proposed a DRL framework to jointly optimize the offloading decisions and resource allocation, with the goal of minimizing the weighted sum of MD's energy consumption and task execution time. The DRL framework utilizes a DNN to learn and improve the offloading policy from the experiences, which completely removes the need of solving hard combinatorial optimization problem.  Besides, we have derived a Gaussian noise-added order-preserving quantization method to efficiently generate offloading actions in the DRL framework. Meanwhile, a low-complexity algorithm has been proposed to accurately evaluate the ETC performance of each generated offloading decision. We have further proposed an one-climb policy to speed up the learning process. Simulation results have demonstrated that the proposed algorithm can achieve near-optimal performance while significantly decreasing the complexity compared to the conventional optimization methods.

\appendices

\section{Proof of Lemma 3.1} \label{appendicesC}
According to \eqref{RT_local}, \eqref{FT_local}, \eqref{RT_edge} and \eqref{FT_edge}, we have
\begin{align}\label{FT1}
FT_{M+1}^{l}&=RT_{M+1}^{l}+\tau_{M+1}^{l}\nonumber\\
&=\max_{k_m\in\textbf{pred($M+1$)}}\left\{(1-a_{k_m})FT_{k_m}^{l}+a_{k_m}(FT_{k_m}^{c}+\tau_{k_m,M+1}^{d})\right\}\nonumber\\
&=\max_{k_m\in\textbf{pred($M+1$)}}\left\{(1-a_{k_m})(RT_{k_m}^{l}+\tau_{k_m}^{l})+a_{k_m}(RT_{k_m}^{c}+\tau_{k_m}^{c}+\tau_{k_m,M+1}^{d})\right\}.
\end{align}

For the term $RT_{k_m}^{l}$ in \eqref{FT1}, we have
\begin{align}\label{RT1}
RT_{k_m}^{l} &= \max_{k_{m-1}\in\textbf{pred($k_m$)}}\left\{(1-a_{k_{m-1}})FT_{k_{m-1}}^{l}+a_{k_{m-1}}(FT_{k_{m-1}}^{c}+\tau_{k_{m-1},k_{m}}^{d})\right\}\nonumber\\
&=\max_{k_{m-1}\in\textbf{pred($k_m$)}}\left\{(1-a_{k_{m-1}})(RT_{k_{m-1}}^{l}+\tau_{k_{m-1}}^{l})+a_{k_{m-1}}(RT_{k_{m-1}}^{c}+\tau_{k_{m-1}}^{c}+\tau_{k_{m-1},k_{m}}^{d})\right\}.
\end{align}

For the term $RT_{k_m}^{c}$ in \eqref{FT1}, we have
\begin{align}\label{RT2}
RT_{k_m}^{c} &= \max_{k_{m-1}\in\textbf{pred($k_m$)}}\left\{(1-a_{k_{m-1}})(FT_{k_{m-1}}^{l}+\tau_{k_{m-1},k_{m}}^{u})+a_{k_{m-1}}FT_{k_{m-1}}^{c}\right\}\nonumber\\
&=\max_{k_{m-1}\in\textbf{pred($k_m$)}}\left\{(1-a_{k_{m-1}})(RT_{k_{m-1}}^{l}+\tau_{k_{m-1}}^{l}+\tau_{k_{m-1},k_{m}}^{u})+a_{k_{m-1}}(RT_{k_{m-1}}^{c}+\tau_{k_{m-1}}^{c})\right\}.
\end{align}

Substituting \eqref{RT1} and \eqref{RT2} into \eqref{FT1}, we have

\begin{align}
FT_{M+1}^{l}=&\max_{k_m\in\textbf{pred($M+1$)}}\left\{(1-a_{k_{m}})\tau_{k_{m}}^{l}+a_{k_{m}}(\tau_{k_{m}}^{c}+\tau_{k_{m},M+1}^{d})\right\}+\max_{k_m\in\textbf{pred($M+1$)}}\max_{k_{m-1}\in\textbf{pred($k_m$)}}\nonumber\\
&\bigg\{(1-a_{k_{m-1}})\tau_{k_{m-1}}^{l}+a_{k_{m-1}}\tau_{k_{m-1}}^{c}+a_{k_{m}}(1-a_{k_{m-1}})\tau_{k_{m-1},k_{m}}^{u}+(1-a_{k_{m}})a_{k_{m-1}}\tau_{k_{m-1},k_{m}}^{d}\bigg\}+\nonumber\\
&\max_{k_m\in\textbf{pred($M+1$)}}\max_{k_{m-1}\in\textbf{pred($k_m$)}}\left\{(1-a_{k_{m-1}})RT_{k_{m-1}}^{l}+a_{k_{m-1}}RT_{k_{m-1}}^{c}\right\}\nonumber\\
=&\max_{k_m\in\textbf{pred($M+1$)}}\left\{(1-a_{k_{m}})\tau_{k_{m}}^{l}+a_{k_{m}}(\tau_{k_{m}}^{c}+\tau_{k_{m},M+1}^{d})\right\}+\max_{k_m\in\textbf{pred($M+1$)}}\max_{k_{m-1}\in\textbf{pred($k_m$)}}\nonumber\\
&\left\{(1-a_{k_{m-1}})\tau_{k_{m-1}}^{l}+a_{k_{m-1}}\tau_{k_{m-1}}^{c}+a_{k_{m}}(1-a_{k_{m-1}})\tau_{k_{m-1},k_{m}}^{u}+(1-a_{k_{m}})a_{k_{m-1}}\tau_{k_{m-1},k_{m}}^{d}\right\}+\nonumber\\
&\max_{k_m\in\textbf{pred($M+1$)}}\max_{k_{m-1}\in\textbf{pred($k_m$)}}\max_{k_{m-2}\in\textbf{pred($k_{m-1}$)}}\bigg\{(1-a_{k_{m-2}})\tau_{k_{m-2}}^{l}+a_{k_{m-2}}\tau_{k_{m-2}}^{c}+\nonumber\\
&a_{k_{m-1}}(1-a_{k_{m-2}})\tau_{k_{m-2},k_{m-1}}^{u}+(1-a_{k_{m-1}})a_{k_{m-2}}\tau_{k_{m-2},k_{m-1}}^{d}\bigg\}+...+\nonumber\\
&\max_{k_m\in\textbf{pred($M+1$)}}\max_{k_{m-1}\in\textbf{pred($k_m$)}}...\max_{k_{1}\in\textbf{pred($k_2$)}}\max_{0\in\textbf{pred($k_1$)}}\left\{a_{k_{1}}\tau_{0,k_{1}}^{u}\right\}\nonumber\\
=&\max\{T_{1},T_{2},...,T_{o},...,T_{O}\},
\end{align}
where $T_o$ is defined in \eqref{T_o}.
%

\section{Proof of Proposition 3.1} \label{appendicesA}
The derivative of $L$ of \eqref{lag} with respect to $\tau_{i}^{l}$ can be expressed as
\begin{align}
\frac{\partial L}{\partial\tau_{i}^{l}}=-\frac{2\kappa\beta_{e}(L_{i})^3}{(\tau_{i}^{l})^3}+\sum_{o\in\Upsilon(i)}\lambda_{o},
\end{align}
where $\frac{\partial L}{\partial\tau_{i}^{l}}$ is a monotonously increasing function with $\tau_{i}^{l}\in[\frac{L_i}{f_{peak}},+\infty)$. Thus, if $\frac{\partial L}{\partial\tau_{i}^{l}}|_{\tau_{i}^{l}=\frac{L_i}{f_{peak}}}>0$, we have $(f_i^l)^*=f_{peak}$. Otherwise, we have
\begin{align}
\tau_{i}^{l}=L_i\sqrt[3]{\frac{2\kappa\beta_{e}}{\sum_{o\in\Upsilon(i)}\lambda_{o}}}\Rightarrow (f_i^l)^*=\frac{L_i}{\tau_{i}^{l}}=\sqrt[3]{\frac{\sum_{o\in\Upsilon(i)}\lambda_{o}^*}{2\kappa\beta_{e}}}.
\end{align}
Hence,
\begin{align}
(f_{i}^{l})^{*}=\min\left\{\sqrt[3]{\frac{\sum_{o\in\Upsilon(i)}\lambda_{o}^*}{2\kappa\beta_{e}}},f_{peak}\right\}.
\end{align}

{\color{blue}{\section{Proof of Corollary 3.1} \label{addappendix}

The derivative of $L$ of (18) with respect to $T_{max}$ can be expressed as
\begin{align}
\frac{\partial L}{\partial T_{max}}=\beta_t-\sum_{o=1}^{O}\lambda_{o}.
\end{align}
By setting $\frac{\partial L}{\partial T_{max}}=0$, we have
\begin{align}
\sum_{o=1}^{O}\lambda_{o}^*=\beta_t.
\end{align}
}}

\section{Optimality Analysis for One-climb Policy} \label{appendicesB}
In the following, we analyze the optimality of the one-climb policy. Suppose that there exists a path $o$ in the task graph, where the optimal offloading decision allows the MD to offload its task data for two times. Under the two-time offloading scheme, for the tasks in $\Psi(o)=\{0,k^{o}_1,...,k^{o}_x,...,k^{o}_{s-1},k^{o}_s,...,k^{o}_n,k^{o}_{n+1},...,k^{o}_y,...,M+1\}$, tasks from $k^{o}_x$ to $k^{o}_{s-1}$ are migrated to the edge server for execution. Then, tasks from $k^{o}_s$ to $k^{o}_n$ prefer local computing, followed by tasks from $k^{o}_{n+1}$ to $k^{o}_y$ migrated to the edge server. We also consider an one-climb scheme for performance comparison, where tasks from $k^{o}_x$ to $k^{o}_{y}$ are executed on the edge server.

We denote the optimal offloading decision and local CPU frequencies in the two-time and one-climb offloading schemes as $\{\hat{\mathbf{a}},\hat{\mathbf{f}}\}$ and $\{\tilde{\mathbf{a}},\tilde{\mathbf{f}}\}$, respectively. By the optimality assumption, we have $\eta(\hat{\mathbf{a}},\hat{\mathbf{f}})<\eta(\tilde{\mathbf{a}},\tilde{\mathbf{f}})$.

For the two-time offloading policy in path $o$, the total execution time from the $k^{o}_x$-th task to the $k^{o}_y$-th task can be expressed as
\begin{align}
\hat{T}_{o}^{k^{o}_x\sim k^{o}_y}=&\sum_{m=x}^{s-1}(\tau_{k^{o}_m}^{c})+\tau_{k^{o}_{s-1},k^{o}_s}^{d}+\sum_{m=s}^{n}(\tau_{k^{o}_m}^{l})+\tau_{k^{o}_n,k^{o}_{n+1}}^{u}+\sum_{m=n+1}^{y}(\tau_{k^{o}_m}^{c}).
\end{align}
As for the one-climb policy in path $o$, we have
\begin{align}
\tilde{T}_{o}^{k^{o}_x\sim k^{o}_y}=\sum_{m=x}^{y}\tau_{k^{o}_m}^{c}.
\end{align}

Since $f^{c}>f_{peak}$, the following inequalities hold for the $k^{o}_s$-th and $k^{o}_n$-th tasks:
\begin{align}
\tau_{k^{o}_s}^{c}<\tau_{k^{o}_s}^{l}<\tau_{k^{o}_s}^{l}+\tau_{k^{o}_{s-1},k^{o}_s}^{d},
\end{align}
\begin{align}
\tau_{k^{o}_n}^{c}<\tau_{k^{o}_n}^{l}<\tau_{k^{o}_n}^{l}+\tau_{k^{o}_n,k^{o}_{n+1}}^{u}.
\end{align}
In addition, we have $\tau_{k^{o}_m}^{c}<\tau_{k^{o}_m}^{l}, m=s,...,n$ for the tasks in the $o$-th path between $k^{o}_s$ and $k^{o}_n$. Therefore, it can be shown that $\hat{T}_{o}^{k^{o}_x\sim k^{o}_y}>\tilde{T}_{o}^{k^{o}_x\sim k^{o}_y}$.

On the other hand, with respect to the energy consumption of the MD from the $k^{o}_x$-th task to the $k^{o}_y$-th task in the $o$-th path, we observe that the two-time offloading scheme consumes more energy compared with the one-climb policy due to the local tasks computing $e_{k^{o}_i}^{l}$ from $k^{o}_s$ to $k^{o}_n$ and the $k^{o}_{n+1}$-th task's offloading $e_{k^{o}_n,k^{o}_{n+1}}^{u}$. That is, $\hat{E}_{o}^{k^{o}_x\sim k^{o}_y}>\tilde{E}_{o}^{k^{o}_x\sim k^{o}_y}$, where $\hat{E}_{o}^{k^{o}_x\sim k^{o}_y}$ and $\tilde{E}_{o}^{k^{o}_x\sim k^{o}_y}$ denote the energy consumption from the $k^{o}_x$-th task to the $k^{o}_y$-th task in the $o$-th path under the two-time and one-climb offloading schemes, respectively.

\begin{figure}
\begin{centering}
\includegraphics[scale=0.6]{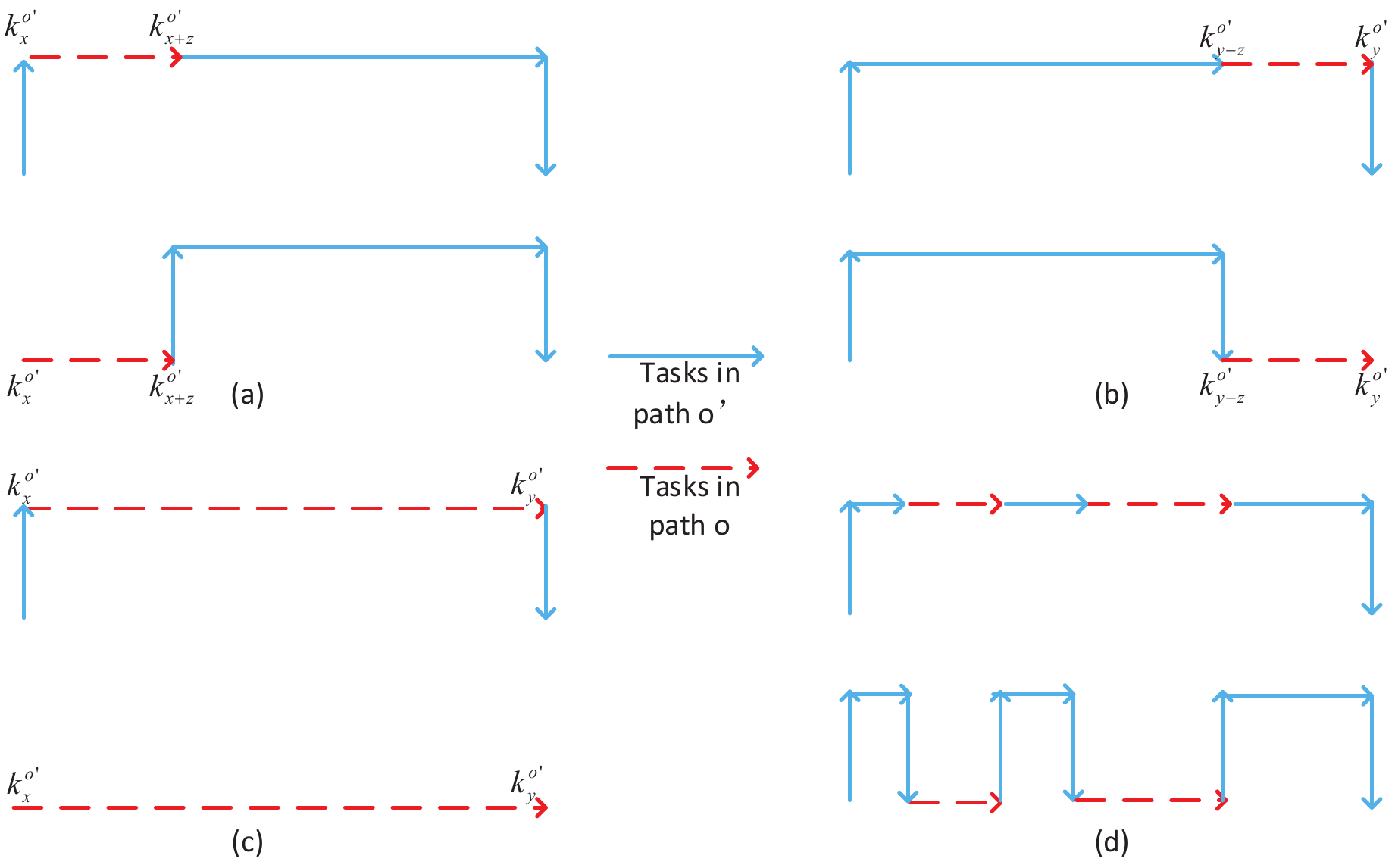}
\vspace{-0.1cm}
 \caption{ Illustration of different offloading decisions at the path $o'$ due to the overlapping tasks belonging to path $o$. }
\end{centering}
\vspace{-0.1cm}
\end{figure}

For another path $o'$ in the task graph $G$, we assume that in the one-climb scheme, tasks from $k^{o'}_x$ to $k^{o'}_y$ are executed on the edge server. Consider the tasks in $\{k^{o}_s,...,k^{o}_n\}$ that the path $o'$ also contains. If $\{k^{o}_s,...,k^{o}_n\}\bigcap\Psi(o')=\emptyset$, we have $\tilde{T}_{o'}=\hat{T}_{o'}$, where $\tilde{T}_{o'}$ is the total execution time in the $o'$-th path under one-climb policy, and $\hat{T}_{o'}$ is the execution time when the tasks in $\{k^{o}_s,...,k^{o}_n\}\bigcap\Psi(o')$ choose to perform local computing due to the two-time offloading scheme in the $o$-th path. Meanwhile, $\tilde{E}_{o'}=\hat{E}_{o'}$, where $\tilde{E}_{o'}$  is the total energy consumption in the $o'$-th path under one-climb policy, and $\hat{E}_{o'}$ is the energy consumption when the tasks in $\{k^{o}_s,...,k^{o}_n\}\bigcap\Psi(o')$ change their offloading decisions due to the two-time offloading scheme in the $o$-th path. Otherwise, if $\{k^{o}_s,...,k^{o}_n\}\bigcap\Psi(o')\neq\emptyset$, we consider the following four cases.
\begin{itemize}
  \item As shown in Fig. 12(a), suppose that the tasks in $\{k^{o}_s,...,k^{o}_n\}$, which the path $o'$ also includes, are the first $z$ tasks offloaded to the edge in path $o'$ under one-climb scheme, i.e., $\{k^{o}_s,...,k^{o}_n\}\bigcap\Psi(o')=\{k^{o'}_x,k^{o'}_{x+1},...,k^{o'}_{x+z}\}$.
 We have
\begin{align}
\hat{T}_{o'}-\tilde{T}_{o'}=\frac{O_{k^{o'}_{x+z},k^{o'}_{x+z+1}}}{R^u(h^u_{k^{o'}_{x+z},k^{o'}_{x+z+1}})}-\frac{O_{k^{o'}_{x-1},k^{o'}_{x}}}{R^u(h^u_{k^{o'}_{x-1},k^{o'}_{x}})}+Y-Z,
\end{align}
and
\begin{align}
\hat{E}_{o'}-\tilde{E}_{o'}=P_{MD}[\frac{O_{k^{o'}_{x+z},k^{o'}_{x+z+1}}}{R^u(h^u_{k^{o'}_{x+z},k^{o'}_{x+z+1}})}-\frac{O_{k^{o'}_{x-1},k^{o'}_{x}}}{R^u(h^u_{k^{o'}_{x-1},k^{o'}_{x}})}]+X,
\end{align}
where $X,Y,Z$ are the total local execution energy consumption, local computing time and edge execution time among the tasks $\{k^{o}_s,...,k^{o}_n\}\bigcap\Psi(o')$ in the path $o'$, respectively. In this case, if $\hat{T}_{o'}>\tilde{T}_{o'}$ and $\hat{E}_{o'}>\tilde{E}_{o'}$ hold, the following inequality needs to be satisfied:
\begin{align}\label{case1}
\Delta^{u}=\frac{O_{k^{o'}_{x+z},k^{o'}_{x+z+1}}}{R^u(h^u_{k^{o'}_{x+z},k^{o'}_{x+z+1}})}-\frac{O_{k^{o'}_{x-1},k^{o'}_{x}}}{R^u(h^u_{k^{o'}_{x-1},k^{o'}_{x}})}<\frac{X+Y-Z}{1+P_{MD}},
\end{align}
where $\Delta^{u}$ denotes the gap of the uplink transmission time associated with two ordered transferred data in $G$. Note that $X+Y$ is a function with respect to the local CPU frequencies $f^l_i, i\in\{k^{o}_s,...,k^{o}_n\}\bigcap\Psi(o')$ and can achieve minimum when $f^l_i=\min\{\sqrt[3]{\frac{1}{2\kappa}},f_{peak}\}, \forall i\in\{k^{o}_s,...,k^{o}_n\}\bigcap\Psi(o')$. Let $(X+Y)^*$
denote the minimum of $X+Y$. Thus, \eqref{case1} can be rewritten as
\begin{align}\label{case4}
\Delta^{u}<\frac{(X+Y)^*-Z}{1+P_{MD}}.
\end{align}

  \item As shown in Fig. 12(b), suppose that the tasks in $\{k^{o}_s,...,k^{o}_n\}$, which also exist in path  $o'$, are the last $z$ tasks offloaded to the edge in path $o'$ under one-climb scheme, i.e., $\{k^{o}_s,...,k^{o}_n\}\bigcap\Psi(o')=\{k^{o'}_{y-z},k^{o'}_{y-z+1},...,k^{o'}_{y}\}$. Similarly, if $\hat{T}_{o'}>\tilde{T}_{o'}$ and $\hat{E}_{o'}>\tilde{E}_{o'}$, we have
\begin{align}\label{case2}
\Delta^{d}=\frac{O_{k^{o'}_{y-z-1},k^{o'}_{y-z}}}{R^d(h^d_{k^{o'}_{y-z-1},k^{o'}_{y-z}})}-\frac{O_{k^{o'}_{y},k^{o'}_{y+1}}}{R^d(h^d_{k^{o'}_{y},k^{o'}_{y+1}})}<X+Y-Z,
\end{align}
where $\Delta^{d}$ denotes the gap of the downlink transmission time associated with two ordered transferred data in $G$. Then, we have
\begin{align}\label{case5}
\Delta^{d}<(X+Y)^*-Z.
\end{align}

  \item  As shown in Fig. 12(c), suppose that the tasks in $\{k^{o}_s,...,k^{o}_n\}$, which the path $o'$ consists of, are the total tasks offloaded to the edge in path $o'$ under one-climb scheme, i.e., $\{k^{o}_s,...,k^{o}_n\}\bigcap\Psi(o')=\{k^{o'}_{x},...,k^{o'}_{y}\}$. If $\hat{T}_{o'}>\tilde{T}_{o'}$ and $\hat{E}_{o'}>\tilde{E}_{o'}$ hold, we have
\begin{align}\label{case3}
\Delta^{ud}=(1+P_{MD})\frac{O_{k^{o'}_{x-1},k^{o'}_{x}}}{R^u(h^u_{k^{o'}_{x-1},k^{o'}_{x}})}+\frac{O_{k^{o'}_{y},k^{o'}_{y+1}}}{R^d(h^d_{k^{o'}_{y},k^{o'}_{y+1}})}<X+Y-Z.
\end{align}
That is,
\begin{align}\label{case6}
\Delta^{ud}<(X+Y)^*-Z.
\end{align}
  \item Otherwise, as shown in Fig. 12(d), we can find that changing the offloading decisions for the tasks $\{k^{o}_s,...,k^{o}_n\}\bigcap\Psi(o')$ from 1 to 0 will lead to multi-time offloading in the path $o'$. According to the above discussion, we have $\hat{T}_{o'}>\tilde{T}_{o'}$ and $\hat{E}_{o'}>\tilde{E}_{o'}$.
\end{itemize}

Overall, if $\hat{T}_{o'}>\tilde{T}_{o'}$ and $\hat{E}_{o'}>\tilde{E}_{o'}$, \eqref{case4}, \eqref{case5} and \eqref{case6} need to hold. Suppose that we have $\hat{T}_{o'}>\tilde{T}_{o'}$ and $\hat{E}_{o'}>\tilde{E}_{o'}$. Then,
%
\begin{align}
\tilde{FT}_{M+1}^{l}(\tilde{\mathbf{a}},\hat{\mathbf{f}})<\hat{FT}_{M+1}^{l}(\hat{\mathbf{a}},\hat{\mathbf{f}}),
\end{align}
where $\tilde{FT}_{M+1}^{l}$ is the total execution time of the task graph $G$ when all the paths follow the one-climb policy, while $\hat{FT}_{M+1}^{l}$ is the final delay when the tasks in path $o$ prefer two-time offloading scheme. Meanwhile,
%
\begin{align}
\tilde{E}(\tilde{\mathbf{a}},\hat{\mathbf{f}})<\hat{E}(\hat{\mathbf{a}},\hat{\mathbf{f}}),
\end{align}
where $\tilde{E}$ denotes the total energy consumption of the task graph $G$ when all the paths follow the one-climb policy, while $\hat{E}$ denotes the total energy consumption when the tasks in path $o$ prefer two-time offloading scheme.

Therefore, we have
\begin{align}
\eta(\hat{\mathbf{a}},\hat{\mathbf{f}})&=\beta_t\hat{FT}_{M+1}^{l}(\hat{\mathbf{a}},\hat{\mathbf{f}})+\beta_e\hat{E}(\hat{\mathbf{a}},\hat{\mathbf{f}})>\beta_t\tilde{FT}_{M+1}^{l}(\tilde{\mathbf{a}},\hat{\mathbf{f}})+\beta_e\tilde{E}(\tilde{\mathbf{a}},\hat{\mathbf{f}})\nonumber\\
&>\beta_t\tilde{FT}_{M+1}^{l}(\tilde{\mathbf{a}},\tilde{\mathbf{f}})+\beta_e\tilde{E}(\tilde{\mathbf{a}},\tilde{\mathbf{f}})=\eta(\tilde{\mathbf{a}},\tilde{\mathbf{f}}),
\end{align}
where the last inequality means that the optimal $\{\hat{\mathbf{f}}\}$ in a two-time offloading scheme is a feasible solution in the one-climb offloading scheme of (P2). Therefore, it contradicts the assumption. To sum up, we have \eqref{case4}, \eqref{case5} and \eqref{case6} if the one-climb policy is optimal.


\begin{footnotesize}
\bibliographystyle{IEEEtran}
\bibliography{IEEEabrv,references}

\begin{thebibliography}{10}
\providecommand{\url}[1]{#1}
\csname url@samestyle\endcsname
\providecommand{\newblock}{\relax}
\providecommand{\bibinfo}[2]{#2}
\providecommand{\BIBentrySTDinterwordspacing}{\spaceskip=0pt\relax}
\providecommand{\BIBentryALTinterwordstretchfactor}{4}
\providecommand{\BIBentryALTinterwordspacing}{\spaceskip=\fontdimen2\font plus
\BIBentryALTinterwordstretchfactor\fontdimen3\font minus
  \fontdimen4\font\relax}
\providecommand{\BIBforeignlanguage}[2]{{%
\expandafter\ifx\csname l@#1\endcsname\relax
\typeout{** WARNING: IEEEtran.bst: No hyphenation pattern has been}%
\typeout{** loaded for the language `#1'. Using the pattern for}%
\typeout{** the default language instead.}%
\else
\language=\csname l@#1\endcsname
\fi
#2}}
\providecommand{\BIBdecl}{\relax}
\BIBdecl

\bibitem{myICC}
J.~Yan, S.~Bi, L.~Huang, and Y.~J. Zhang, ``Deep reinforcement learning based
  offloading for mobile edge computing with general task graph,''
  \emph{{submitted to} IEEE ICC}, Jun. 2020.

\bibitem{MECsurvey1}
Y.~Mao, C.~You, J.~Zhang, K.~Huang, and K.~B. Letaief, ``A survey on mobile
  edge computing: The communication perspective,'' \emph{IEEE Commun. Surveys
  Tuts.}, vol.~19, no.~4, pp. 2322--2358, Fourthquarter 2017.

\bibitem{MECsurvey2}
W.~Shi, J.~Cao, Q.~Zhang, Y.~Li, and L.~Xu, ``Edge computing: Vision and
  challenges,'' \emph{IEEE Internet Things J.}, vol.~3, no.~5, pp. 637--646,
  Oct. 2016.

\bibitem{MEC3}
S.~Bi and Y.~J. Zhang, ``Computation rate maximization for wireless powered
  mobile-edge computing with binary computation offloading,'' \emph{{IEEE}
  Trans. Wireless Commun.}, vol.~17, no.~6, pp. 4177--4190, Jun. 2018.

\bibitem{xu}
F.~Wang, J.~Xu, X.~Wang, and S.~Cui, ``Joint offloading and computing
  optimization in wireless powered mobile-edge computing systems,''
  \emph{{IEEE} Trans. Wireless Commun.}, vol.~17, no.~3, pp. 1784--1797, Mar.
  2018.

\bibitem{MEC2}
C.~You, K.~Huang, and H.~Chae, ``Energy efficient mobile cloud computing
  powered by wireless energy transfer,'' \emph{{IEEE} J. Sel. Areas Commun.},
  vol.~34, no.~5, pp. 1757--1771, May 2016.

\bibitem{MEC5}
W.~Zhang, Y.~Wen, K.~Guan, D.~Kilper, H.~Luo, and D.~O. Wu, ``Energy-optimal
  mobile cloud computing under stochastic wireless channel,'' \emph{{IEEE}
  Trans. Wireless Commun.}, vol.~12, no.~9, pp. 4569--4581, Sept. 2013.

\bibitem{MEC7}
M.~H. Chen, B.~Liang, and M.~Dong, ``Joint offloading decision and resource
  allocation for multi-user multi-task mobile cloud,'' in \emph{Proc. IEEE
  ICC}, May 2016.

\bibitem{MEC4}
T.~Q. Dinh, J.~Tang, Q.~D. La, and T.~Q.~S. Quek, ``Offloading in mobile edge
  computing: Task allocation and computational frequency scaling,''
  \emph{{IEEE} Trans. Commun.}, vol.~65, no.~8, pp. 3571--3584, Aug. 2017.

\bibitem{partial1}
C.~{You}, K.~{Huang}, H.~{Chae}, and B.~{Kim}, ``Energy-efficient resource
  allocation for mobile-edge computation offloading,'' \emph{{IEEE} Trans.
  Wireless Commun.}, vol.~16, no.~3, pp. 1397--1411, March 2017.

\bibitem{partial2}
Y.~{Wang}, M.~{Sheng}, X.~{Wang}, L.~{Wang}, and J.~{Li}, ``Mobile-edge
  computing: Partial computation offloading using dynamic voltage scaling,''
  \emph{{IEEE} Trans. Commun.}, vol.~64, no.~10, pp. 4268--4282, Oct 2016.

\bibitem{cs_taskgraph}
Y.-K. Kwok and I.~Ahmad, ``Dynamic critical-path scheduling: an effective
  technique for allocating task graphs to multiprocessors,'' \emph{{IEEE}
  Trans. Parallel Distrib. Syst.}, vol.~7, no.~5, pp. 506--521, May 1996.

\bibitem{single4}
W.~Zhang, Y.~Wen, and D.~O. Wu, ``Collaborative task execution in mobile cloud
  computing under a stochastic wireless channel,'' \emph{{IEEE} Trans. Wireless
  Commun.}, vol.~14, no.~1, pp. 81--93, Jan. 2015.

\bibitem{single5}
W.~Zhang and Y.~Wen, ``Energy-efficient task execution for application as a
  general topology in mobile cloud computing,'' \emph{{to appear in} IEEE
  Transactions on Cloud Computing}.

\bibitem{single6}
S.~E. {Mahmoodi}, R.~N. {Uma}, and K.~P. {Subbalakshmi}, ``Optimal joint
  scheduling and cloud offloading for mobile applications,'' \emph{IEEE
  Transactions on Cloud Computing}, vol.~7, no.~2, pp. 301--313, April 2019.

\bibitem{single7}
C.~{Tang}, X.~{Wei}, S.~{Xiao}, W.~{Chen}, W.~{Fang}, W.~{Zhang}, and M.~{Hao},
  ``A mobile cloud based scheduling strategy for industrial internet of
  things,'' \emph{IEEE Access}, vol.~6, pp. 7262--7275, 2018.

\bibitem{multi1}
S.~Guo, B.~Xiao, Y.~Yang, and Y.~Yang, ``Energy-efficient dynamic offloading
  and resource scheduling in mobile cloud computing,'' in \emph{Proc. IEEE
  INFOCOM}, Apr. 2016.

\bibitem{mypaper}
J.~{Yan}, S.~{Bi}, Y.~J. {Zhang}, and M.~{Tao}, ``Optimal task offloading and
  resource allocation in mobile-edge computing with inter-user task
  dependency,'' \emph{{IEEE} Trans. Wireless Commun.}, vol.~19, no.~1, pp.
  235--250, Jan 2020.

\bibitem{DRL2}
M.~{Min}, L.~{Xiao}, Y.~{Chen}, P.~{Cheng}, D.~{Wu}, and W.~{Zhuang},
  ``Learning-based computation offloading for iot devices with energy
  harvesting,'' \emph{IEEE Transactions on Vehicular Technology}, vol.~68,
  no.~2, pp. 1930--1941, Feb 2019.

\bibitem{DRL1}
X.~{Chen}, H.~{Zhang}, C.~{Wu}, S.~{Mao}, Y.~{Ji}, and M.~{Bennis}, ``Optimized
  computation offloading performance in virtual edge computing systems via deep
  reinforcement learning,'' \emph{IEEE Internet of Things Journal}, pp. 1--1,
  2019.

\bibitem{DRL4}
L.~{Huang}, S.~{Bi}, and Y.~J. {Zhang}, ``Deep reinforcement learning for
  online computation offloading in wireless powered mobile-edge computing
  networks,'' \emph{{IEEE} Trans. Mobile Comput.}, pp. 1--1, 2019.

\bibitem{DRL5}
J.~{Wang}, J.~{Hu}, G.~{Min}, W.~{Zhan}, Q.~{Ni}, and N.~{Georgalas},
  ``Computation offloading in multi-access edge computing using a deep
  sequential model based on reinforcement learning,'' \emph{IEEE Communications
  Magazine}, vol.~57, no.~5, pp. 64--69, May 2019.

\bibitem{add1}
V.~M. et~al., ``Asynchronous methods for deep reinforcement learning,'' in
  \emph{Proc. 33rd Int. Conf. Mach. Learn.}, New York, NY, USA, Jun. 2016, pp.
  1928--1937.

\bibitem{add3}
K.~D. Vogeleer, G.~Memmi, P.~Jouvelot, and F.~Coelho, ``The energy/frequency
  convexity rule: Modeling and experimental validation on mobile devices,'' in
  \emph{Proc. Int. Conf. Parallel Process. Appl. Math. (PPAM)}, Warsaw, Poland,
  Sep. 2013, pp. 793--803.

\bibitem{add2}
T.~D. Burd and R.~W. Broderson, ``Processor design for portable systems,'' in
  \emph{J. VLSI Signal Process. Syst.}, vol.~13, Aug./Sep. 1996, pp. 203--221.

\bibitem{add4}
M.~Ehrgott, \emph{Multicriteria Optimization}.\hskip 1em plus 0.5em minus
  0.4em\relax New York, NY, USA: Springer, 2006.

\bibitem{convex}
S.~Boyd and L.~Vandenberghe, \emph{\normalfont{Convex Optimization}}.\hskip 1em
  plus 0.5em minus 0.4em\relax Cambidge University Press, 2004.

\bibitem{add5}
E.~{Perahia} and D.~C. {Cox}, ``Shadow fading correlation between uplink and
  downlink,'' in \emph{IEEE VTS 53rd Vehicular Technology Conference, Spring
  2001. Proceedings (Cat. No.01CH37202)}, vol.~1, May 2001, pp. 308--312 vol.1.

\bibitem{GS2}
L.~P. Qian, Y.~J.~A. Zhang, and M.~Chiang, ``Distributed nonconvex power
  control using gibbs sampling,'' \emph{{IEEE} Trans. Commun.}, vol.~60,
  no.~12, pp. 3886--3898, Dec 2012.

\end{thebibliography}
\end{footnotesize}

%
%
%
%

\end{document}